\documentclass[iop]{emulateapj}

\newcommand{\sqcm}{${\rm cm^{-2}}$}

\newcommand{\mum}{${\rm \mu m}$}

\newcommand{\ammonium}{${\rm NH_{4}^+}$}

\slugcomment{Accepted for publication in The Astrophysical Journal, January 06, 2011}

\shorttitle{Ice and Dust in Isolated Dense Cores}

\shortauthors{Boogert, et al.}

\begin{document}

\title{Ice and Dust in the Quiescent Medium of Isolated Dense
  Cores\altaffilmark{1}}

\author{A. C. A. Boogert\altaffilmark{2},
        T. L. Huard\altaffilmark{3,4},
        A. M. Cook\altaffilmark{5,6},
        J. E. Chiar\altaffilmark{7},
        C. Knez\altaffilmark{3,8},
        L. Decin\altaffilmark{9,10},
        G. A. Blake\altaffilmark{11},
        A. G. G. M. Tielens\altaffilmark{12},
        E. F. van Dishoeck\altaffilmark{12, 13}}

\altaffiltext{1}{Some of the data presented herein were obtained at
    the W.M. Keck Observatory, which is operated as a scientific
    partnership among the California Institute of Technology, the
    University of California and the National Aeronautics and Space
    Administration. The Observatory was made possible by the generous
    financial support of the W.M. Keck Foundation.}
\altaffiltext{2}{IPAC, NASA Herschel Science Center,
  Mail Code 100-22, California Institute of Technology, Pasadena, CA
  91125, USA (email: aboogert@ipac.caltech.edu)}
\altaffiltext{3}{Department of Astronomy, University of Maryland,
  College Park, MD 20742, USA}
\altaffiltext{4}{Columbia Astrophysics Laboratory, Columbia
  University, New York, NY 10027, USA}
\altaffiltext{5}{New York Center for Astrobiology and Department of
  Physics, Applied Physics \& Astronomy, Rensselaer Polytechnic
  Institute, 110 Eighth Street, Troy, NY 12180, USA}
\altaffiltext{6}{IPAC, Mail Code 100-22, California Institute of
  Technology, Pasadena, CA 91125, USA}
\altaffiltext{7}{SETI Institute, Carl Sagan Center, 189 Bernardo Av.,
  Mountain View, CA 94043, USA}
\altaffiltext{8}{Johns Hopkins University Applied Physics Laboratory,
  11100 Johns Hopkins Road, Laurel, MD 20723, USA}
\altaffiltext{9}{Instituut voor Sterrenkunde, Katholieke Universiteit
  Leuven, Celestijnenlaan 200D, 3001 Leuven, Belgium}
\altaffiltext{10}{Sterrenkundig Instituut Anton Pannekoek, University
  of Amsterdam, Science Park 904, NL-1098 Amsterdam, The Netherlands}
\altaffiltext{11}{Division of GPS, Mail Code 150-21, California
  Institute of Technology, Pasadena, CA 91125, USA}
\altaffiltext{12}{Leiden Observatory, Leiden University, PO Box 9513,
  2300 RA Leiden, the Netherlands}
\altaffiltext{13}{Max Planck Institut f\"ur Extraterrestrische Physik
  (MPE), Giessenbachstr. 1, 85748 Garching, Germany}

\begin{abstract}

The relation between ices in the envelopes and disks surrounding YSOs
and those in the quiescent interstellar medium is investigated.  For a
sample of 31 stars behind isolated dense cores, ground-based and {\it
  Spitzer} spectra and photometry in the 1-25 \mum\ wavelength range
are combined.  The baseline for the broad and overlapping ice features
is modeled, using calculated spectra of giants, H$_2$O ice and
silicates. The adopted extinction curve is derived empirically. Its
high resolution allows for the separation of continuum and feature
extinction. The extinction between 13-25 \mum\ is $\sim$50\% relative
to that at 2.2 \mum. The strengths of the 6.0 and 6.85
\mum\ absorption bands are in line with those of YSOs.  Thus, their
carriers, which, besides H$_2$O and CH$_3$OH, may include NH$_4^+$,
HCOOH, H$_2$CO and NH$_3$, are readily formed in the dense core phase,
before stars form.  The 3.53 \mum\ C-H stretching mode of solid
CH$_3$OH was discovered.  The CH$_3$OH/H$_2$O abundance ratios of
5-12\% are larger than upper limits in the Taurus molecular cloud. The
initial ice composition, before star formation occurs, therefore
depends on the environment. Signs of thermal and energetic processing
that were found toward some YSOs are absent in the ices toward
background stars. Finally, the peak optical depth of the 9.7
\mum\ band of silicates relative to the continuum extinction at 2.2
\mum\ is significantly shallower than in the diffuse interstellar
medium. This extends the results of \citet{chi07} to a larger sample
and higher extinctions.

\end{abstract}


\section{Introduction}~\label{sec:intro}

Planets and comets are assembled from the gas, dust and ices in
circumstellar disks that in turn evolve from the envelopes and clouds
surrounding YSOs.  Many key questions about the molecular composition
of protostellar environments are still unanswered.  The infrared
spectra of YSOs are rich in absorption features caused by the
vibrational transitions of silicates and of solid H$_2$O, CO,
$^{13}$CO, CH$_3$OH, NH$_3$, OCS, CH$_4$, CO$_2$, $^{13}$CO$_2$, and
possibly of OCN$^-$, NH$_4^+$, HCOOH and HCOO$^-$ \citep{gib04, boo04,
  boo08, obe08, zas09, bot10}.  Band profile studies revealed that both CO
and CO$_2$ are present in multiple components of mixed or pure ices
that are formed or destroyed at different stages of the cloud and
envelope evolution \citep{tie91, ger99, boo02, pon03a, pon08}.  It was
also shown that the prominent 6.0 and 6.85 \mum\ absorption features
toward low and high mass YSOs are caused by the overlapping modes of
simple species: H$_2$O, H$_2$CO, HCOOH, NH$_3$, NH$_4^+$, and CH$_3$OH
\citep{kea01, boo08}. Not all absorption observed in this wavelength
region is explained, however, such as the very broad component labeled
``C5'' in \citet{boo08}.  Ices processed by heat, UV photons or cosmic
ray hits are potential candidates. Both ions \citep{sch03} and
hydrocarbons \citep{gib02} have absorption features that are spread
over much of the 5-8~\mum\ region. PAH molecules embedded in the ices
could be responsible for some of the absorption as well.

These molecules are formed by a complex interplay between numerous
formation and destruction processes.  Grain surface chemistry likely
dominates the formation of molecules (e.g., H$_2$O, CH$_4$, CO$_2$) in
cold, dense molecular clouds, leading to icy mantles (e.g.,
\citealt{tie82}).  In the protostellar environment these ices may
sublimate through the influence of shocks or photons in the inner
envelopes and disk surfaces.  The formation of complex molecules in
the ices may be triggered by heating and by the absorption of
energetic photons \citep{sch93, gre95} originating in the protostar-disk
boundary layer.  In the most shielded areas (outer envelopes, disk
mid-planes) deeply penetrating cosmic rays can induce similar effects.
It is still quite uncertain which molecule formation and destruction
processes are important in these different environments.  What is the
composition of ices in pristine clouds, prior to star formation?  To
what degree does the solar system molecular inventory resemble
quiescent cloud inventories?

A rigorous observational approach is required to further constrain the
origin of unidentified absorption bands, especially in the
5-8~\mum\ region, and to disentangle the many effects of the physical
environment on molecular evolution.  An established technique is to
contrast the molecular composition toward YSOs with that toward
background stars tracing quiescent molecular cloud material (e.g.,
\citealt{whi83, whi98}).  Large abundances of the most volatile
species (e.g., pure, ``apolar'' CO ice) are observed in quiescent
lines of sight, which is attributed to the absence of nearby heating
sources. Likewise, amorphous ice structures are observed in these
sight-lines, whereas toward YSOs they are sometimes crystalline
\citep{smi89, ger99, pon08}. Tight upper limits have been set to the
abundance of the species OCN$^-$ (cyanate; \citealt{whi01}) and
CH$_3$OH \citep{chi95} toward background stars, suggesting that
specific physical conditions found only around YSOs may be needed to
initiate their formation.

So far, mid-infrared ($\lambda>5$ \mum) studies have focused on only a
few background stars: four toward the Taurus molecular cloud, one
toward Serpens, and one toward IC~5146.  \citet{kne05} show that the
absorption bands in the 5-8~\mum\ wavelength region are similar to
those toward YSOs. The 15~\mum\ CO$_2$ band shows a non-polar ice
component consistent with the pristine nature of the Taurus cloud
\citep{ber05}, a conclusion that can be drawn for Serpens and IC 5146
as well \citep{whi09}.

The studied samples are too small to fully assess the ice inventory
prior to star formation, given the wide range of environmental
conditions possible.  Much larger samples of background stars are
needed, covering wider ranges of extinctions and cloud types.
Selecting these samples is now possible using the catalogs produced by
wide field {\it Spitzer} IRAC and MIPS mapping projects, in
combination with the Two Micron All Sky Survey (2MASS;
\citealt{skr06}).  In this work, background stars are selected from
the catalogs of nearby ($<$350 pc), isolated, compact ($<5$ arcmin),
dense ($>10^4\ {\rm cm^{-3}}$) cores of the {\it Spitzer} Legacy team
``From Molecular Cores to Planet Forming Disks'' (c2d;
\citealt{eva03}).  These cores were originally identified in optically
selected regions of extinction by observing transitions of NH$_3$
sensitive to high densities ($\geq 10^3\ {\rm cm}^{-3}$;
\citealt{mye83}).  Based on IRAS colors, a subdivision was made
between starless cores and cores containing stars, and it was
concluded that roughly half the cores contain stars \citep{bei86}.
With the sensitive {\it Spitzer} MIPS and IRAC cameras, deep searches
revealed new YSOs, down to stellar masses of 0.01 $M_\odot$, even in
some cores that were previously categorized as ``starless''
\citep{you04}.

Studying the interstellar medium in isolated cores is attractive,
because it lacks the environmental influences of outflows and the
resulting turbulence often thought to dominate in regions of clustered
star formation (e.g., $\rho$ Oph, Serpens).  Their star formation time
scales can be significantly larger due to the dominance of magnetic
fields over turbulence and the resulting slow process of ambipolar
diffusion \citep{shu87, eva99}. The longer time scales may alter the
gas composition, such as reduced C/CO and H/H$_2$ ratios, which would
qualitatively alter the grain surface chemistry.  Furthermore, the
longer exposure time of the ices to cosmic rays or cosmic ray induced
ultraviolet photons may initiate reactions with large energy barriers
leading to the formation of complex species \citep{gre95, ber02}.

While ices in the quiescent medium toward isolated dense cores have
rarely been studied, the 9.7~\mum\ band of silicates received more
attention.  \citet{chi07} found that toward the IC 5146, B68 and B59
cores, as well as some lines of sight through the Serpens and Taurus
molecular clouds, the peak optical depth of the 9.7~\mum\ band
relative to the K-band extinction is significantly shallower compared
to the diffuse medium.  Here, this relation will be revisited with a
much larger sample. With a simultaneous analysis of the ices bands,
some possible explanations of this relation can be addressed, e.g.,
the effects of grain growth.

Also, in the process of disentangling ice and dust features from the
attenuated stellar spectra, an extinction curve needs to be assumed.
Broad-band and spectroscopic studies have shown that the extinction
remains high even at 25~\mum\ \citep{cha09, mcc09}. In this work, this
result will be re-assessed, using {\it Spitzer} spectra and
source-specific models. In the process a high resolution extinction
curve will be produced, in which absorption features and continuum
extinction can be separated.

Finally, this study of isolated dense cores is part of a larger
program to investigate dust and ices in a range of environments. While
here many cores are observed with a few background stars per core,
\citet{chi11} study many sight-lines behind one specific core in the
IC~5146 region in much detail. Efforts are under way to study large
nearby clouds as well.

The selection of the background stars is described in \S\ref{sec:sou},
and the reduction of the ground-based and {\it Spitzer} spectra in
\S\ref{sec:obs}. In \S\ref{sec:cont} the procedure to fit the stellar
continua is presented. This is a crucial step in the analysis. For
this, a ``high resolution'' extinction curve is needed in which ice
and silicate features can be separated from continuum extinction.
This extinction curve is derived in \S\ref{sec:ext}. Subsequently, in
\S\ref{sec:col}, the peak and integrated optical depths of the ice and
dust features are derived, as well as column densities for the known
identifications. Then in \S\ref{sec:60}, the derived parameters are
correlated with each other and with previously studied YSOs.  In
\S\ref{sec:dis1}, the comparison of ice abundances in the background
stars and YSOs is discussed, followed by a discussion on observational
tracers of ice processing in \S\ref{sec:dis2}.  \S\ref{sec:dis3} shows
how the observations of the ices further constrain explanations for
the deviating extinction curve and $A_{\rm K}$ versus $\tau_{9.7}$
relation in dense clouds. Finally, the conclusions are summarized and
an outlook to future studies is presented in \S\ref{sec:concl}.

\begin{deluxetable*}{llrll}[b!]
\tabletypesize{\scriptsize}
\tablecolumns{5}
\tablewidth{0pc}
\tablecaption{Source Sample~\label{t:sample}}
\tablehead{
\colhead{Source}& \colhead{Core\tablenotemark{a}} & \colhead{AOR key\tablenotemark{b}} & \colhead{Module\tablenotemark{c}} & \colhead{$\lambda_{\rm NIR}$\tablenotemark{d}}\\
\colhead{2MASS~J}& \colhead{   }   & \colhead{ } & \colhead{ } & \colhead{\mum}\\}
\startdata
04215402$+$1530299                  & IRAM 04191*      & 0014900992                   & SL                        & 2.82-4.14 \\
08052135$-$3909304                  & BHR 16           & 0014896128                   & SL, LL2                   & \nodata   \\
08093135$-$3604035                  & CG 30-31*        & 0014898432                   & SL, LL2                   & \nodata   \\
08093468$-$3605266                  & CG 30-31*        & 0014898176                   & SL                        & \nodata   \\
12014301$-$6508422                  & DC 297.7-2.8*    & 0014907136                   & SL, LL2                   & \nodata   \\
12014598$-$6508586                  & DC 297.7-2.8*    & 0014907392                   & SL, LL2                   & \nodata   \\
15421547$-$5248146                  & DC 3272+18       & 0014899456                   & SL                        & \nodata   \\
15421699$-$5247439                  & DC 3272+18       & 0014899968\tablenotemark{e}  & SL                        & \nodata   \\
17111501$-$2726180                  & B 59*            & 0014900480                   & SL                        & 2.41-4.14 \\
17111538$-$2727144                  & B 59*            & 0014895104                   & SL                        & 2.41-4.14 \\
17112005$-$2727131                  & B 59*            & 0014894848                   & SL                        & 2.38-4.14 \\
17155573$-$2055312                  & L 100*           & 0014901760                   & SL                        & 2.38-4.14 \\
17160467$-$2057072                  & L 100*           & 0014902016                   & SL                        & 2.39-4.14 \\
17160860$-$2058142                  & L 100*           & 0014901248                   & SL                        & 2.38-4.14 \\
18140712$-$0708413                  & L 438            & 0014905088                   & SL                        & 2.82-4.14 \\
18160600$-$0225539                  & CB 130-3         & 0014896897\tablenotemark{f}  & SL                        & 2.82-4.14 \\
18165296$-$1801287                  & L 328            & 0014908928                   & SL                        & 2.41-4.14 \\
18165917$-$1801158                  & L 328            & 0014907648                   & SL, LL2\tablenotemark{g}  & 2.41-4.14 \\
18170426$-$1802408                  & L 328            & 0014907904                   & SL, LL2\tablenotemark{g}  & 2.06-4.14 \\
18170429$-$1802540                  & L 328            & 0014904064                   & SL, LL2                   & 2.39-4.09 \\
18170470$-$0814495                  & L 429-C          & 0014908160                   & SL                        & 2.38-4.14 \\
18170957$-$0814136                  & L 429-C          & 0014898944                   & SL, LL2                   & 2.38-4.14 \\
18171181$-$0814012                  & L 429-C          & 0014908416                   & SL                        & 2.06-4.14 \\
18171366$-$0813188                  & L 429-C          & 0014908672                   & SL                        & 2.06-4.14 \\
18172690$-$0438406                  & L 483*           & 0014905600                   & SL, LL2                   & 2.38-4.14 \\
19201597$+$1135146                  & CB 188*          & 0014897408                   & SL                        & 2.41-4.14 \\
19201622$+$1136292                  & CB 188*          & 0014897664                   & SL, LL2                   & 2.06-4.14 \\
19214480$+$1121203                  & L 673-7          & 0014904576                   & SL                        & 2.06-4.14 \\
21240517$+$4959100                  & L 1014           & 0014902528                   & SL, LL2                   & 1.49-4.14 \\
21240614$+$4958310                  & L 1014           & 0014909696                   & SL                        & 1.49-4.14 \\
22063773$+$5904520                  & L 1165*          & 0014903040                   & SL                        & 1.49-4.14 \\
                                    &                  &                              &                           &           \\
04393886$+$2611266\tablenotemark{h} & Taurus MC        & 0005637632                   & SL, SH                    & 2.59-5.31\tablenotemark{j} \\
18300061$+$0115201\tablenotemark{i} & Serpens MC       & 0011828224                   & SL, SH                    & 2.82-4.14 \\
\enddata
\tablenotetext{a}{Star-forming cores as defined by IRAS studies are indicated with a ``*''.}
\tablenotetext{b}{Identification number for {\it Spitzer} observations}
\tablenotetext{c}{{\it Spitzer}/IRS modules used: SL=Short-Low (5-14~\mum, $R\sim100$), LL2=Long-Low 2 (14-21.3~\mum, $R\sim100$), SH=Short-High (10-20~\mum, $R\sim600$)}
\tablenotetext{d}{Wavelength coverage of complementary near-infrared ground-based observations, excluding the ranges $\sim$1.79-2.06 and  $\sim$2.55-2.82~\mum\ blocked by the Earth's atmosphere.}
\tablenotetext{e}{Averaged with AOR key 0014899712}
\tablenotetext{f}{Also observed in AOR key 0014896896, but rejected due to high solar activity}
\tablenotetext{g}{~LL2 observed, but not included in the analysis because of source confusion}
\tablenotetext{h}{Well-studied background star Elia 3-16 (\citealt{kne05} and references therein)}
\tablenotetext{i}{Well-studied background star [EC92] 118 (\citealt{kne05} and references therein)}
\tablenotetext{j}{Near-infrared data obtained from the ISO archive \citep{whi98}}
\end{deluxetable*}

\section{Source Selection}~\label{sec:sou}

Background stars were selected from a sample of 69 isolated molecular
cloud cores that were mapped with {\it Spitzer}/IRAC and MIPS by the
c2d Legacy team \citep{eva03, eva07}.  The cores are limited by size
($<5$ arcmin), distance ($<350$ pc), and density ($>10^4$ cm$^{-3}$),
and both historically designated starless and star-forming cores (from
IRAS colors; \citealt{bei86}) are included.  The catalogs generated by
the c2d team were systematically searched for sources meeting the
following criteria:

\begin{enumerate}

\item The overall SED (2MASS 1-2~\mum, IRAC 3-8~\mum, MIPS 24~\mum)
  must be that of a reddened Rayleigh-Jeans curve.

\item Regions of high extinction ($A_V \ge\ 5$ mag), associated with
  the core, must be traced. The regional extinction was measured using
  the Near-Infrared Color Excess ({\it{NICE}}) method
  (e.g., \citealt{lad99}), which uses five 2MASS sources nearest each
  target to derive statistically reliable extinctions.

\item Fluxes must be high enough ($>$5~mJy at 8.0~\mum\ or $>$30~mJy
  at 15~\mum) to obtain {\it Spitzer}/IRS spectra of high quality
  (S/N$>$100) within $\sim$30 minutes of observing time per
  module. This is needed to detect the often weak ice absorption
  features and determine their shapes and peak positions.

\end{enumerate}

Thus, a sample of 32 candidate background stars was selected.  One of
those (2MASS J17111631-2725144 behind B59) will not be analyzed here,
as its IRS spectrum shows a silicate emission band and no ice bands.
The remaining 31 sources (Table~\ref{t:sample}) span a range of
$A_{\rm K}$ values and are located behind 16 isolated cores. Eight out
of sixteen cores are ``starless'' in the IRAS definition, although
some of those have {\it Spitzer}-detected YSO candidates (e.g., L1014;
\citealt{you04}). The selected background stars are usually within
$\sim$60$''$ of these YSO candidates. {\it Spitzer} spectra of many
YSO candidates were obtained, but they will be presented in a
forthcoming paper.

Finally, to compare the analysis methods presented here with previous
work \citep{kne05}, two well-studied stars behind the Taurus and
Serpens molecular clouds were included in the sample: 2MASS
J04393886+2611266 (Elia~3-16) and 2MASS J18300061+0115201 (CK~2,
[EC92]~118).

\section{Observations and Data Reduction}~\label{sec:obs}

{\it Spitzer}/IRS spectra of background stars toward isolated dense
cores were obtained as part of a dedicated Open Time program (PID
20604). The complementary Taurus and Serpens background stars were
observed in the c2d program (PID 172).  Table~\ref{t:sample} lists all
sources with their AOR keys, and the IRS modules they were observed
in. The spectra were extracted and calibrated from the two-dimensional
``BCD'' spectra produced by the standard {\it Spitzer} pipeline
(version S15.3.0), using exactly the same method and routines
discussed in \citet{boo08}. Uncertainties (1$\sigma$) for each
spectral point were calculated using the ``func'' frames provided by
the {\it Spitzer} pipeline.


The {\it Spitzer} spectra were complemented by ground-based Keck/NIRSPEC
\citep{mcl98} H, K and L-band spectra at a resolving power of
$R$=2000. These were reduced in a way standard for ground-based
long-slit spectra, using bright, nearby main sequence stars as
telluric and photometric standards.  In the end, all spectra were
multiplied along the flux scale in order to match near-infrared
broad-band photometry from 2MASS \citep{skr06} as well as {\it Spitzer} IRAC
and MIPS photometry from the c2d catalogs \citep{eva07}, using the
appropriate filter profiles.  The same photometry is used in the
continuum determination discussed in \S\ref{sec:res}.  Catalog flags
were taken into account, such that the photometry of sources listed as
being confused within a 2\arcsec\ radius or being located within
2\arcsec\ of a mosaic edge were treated as upper limits. The c2d
catalogs do not include flags for saturation. Therefore, photometry
exceeding the IRAC saturation limits (at the appropriate integration
times) were flagged as lower limits. Finally, since the 2MASS and
{\it Spitzer} photometry represent different surveys, and their relative
calibration is very important for this work, the uncertainties in the
{\it Spitzer} photometry were increased by adding the zero-point magnitude
uncertainties listed in Table 21 of \citet{eva07}.

\section{Results}~\label{sec:res}

The observed spectra (left panels of Fig.~\ref{f:obs1}) show many
distinct absorption features on top of reddened stellar continua: 3.0,
6.0, 6.8, 9.7, and 15~\mum. These are attributed to ices and
silicates, and were previously observed toward many YSOs (e.g.,
\citealt{boo08, pon08}) and some background stars (e.g.,
\citealt{kne05, ber05}). At a lower level, weak features from the
stellar atmosphere are present as well (e.g., 2.4 and 8.0~\mum).  The
separation of interstellar and photospheric features is essential for
this work and is discussed next.

\subsection{Continuum Determination}~\label{sec:cont}

The continua and some of the absorption features in the background
star spectra were fitted by minimizing the reduced-$\chi ^2$ of a
model consisting of the following parameters:

\begin{itemize}

\item {\it Spectral type.} An accurate spectral type determination is
  necessary to separate interstellar features and features from the
  stellar atmosphere, as well as to know the overall spectral energy
  distribution. Correction of the stellar CO and SiO bands at 5.3 and
  8.0~\mum\ is particularly important, as they blend with the
  6.0~\mum\ ice band and the 9.7~\mum\ band of silicates.
  One-dimensional, opacity-sampled spectra were calculated using the
  MARCS 2 model atmospheres following \citet{dec04}.  Only giants
  (luminosity class III) were considered (Table~\ref{t:spt}), as these
  are bright, and most likely to have made it into the source
  selection.  The blend of CO overtone lines at 2.25-2.60~\mum\ is a
  sensitive tracer of spectral type. For the later spectral types
  ($>$M0~III), the 4.7~\mum\ fundamental CO transitions become
  prominent as well. Although most of this region was not observed,
  its edge at 4.0-4.1~\mum\ was; it drops off sharply for spectral
  types $>$M0~III and was used as another independent indicator of
  spectral type. For the earliest spectral types (late G), the
  2.16~\mum\ H-I Br$\gamma$ line is important.  Finally, because the
  opacity-sampled model spectral lines can only be compared directly
  to low resolution observations, the observations and models were
  smoothed to a resolving power of $R$=100 before fitting them.

\item {\it Continuum extinction.} The goal of this work is to measure
  the strength and shape of dust and ice absorption features, and
  therefore a continuum-only extinction curve should be applied to the
  stellar models. Published extinction curves are usually based on
  broad-band photometry (e.g., \citealt{ind05, cha09}), including
  continuum and feature extinction and applying large interpolations
  between, e.g., the IRAC 8~\mum\ and MIPS
  24~\mum\ photometry. Therefore, an independent ``high resolution''
  extinction curve is derived (\S\ref{sec:ext}), from which a
  continuum-only extinction curve was extracted (dashed red line in
  Fig.~\ref{f:ext}b). The latter was applied to the stellar models
  with $A_{\rm K}$ as a free parameter.

\item {\it H$_2$O ice column density.} Broad H$_2$O ice features cover
  much of the observed background star spectra (Fig.~\ref{f:ext}). For
  sources with observed $L$-band spectra, the strong, relatively sharp
  3.0~\mum\ band (O-H stretch mode) constrains the H$_2$O ice column
  density well. This significantly improves the reliability of the
  overall fits, because in the 5-8 and 11-20~\mum\ ranges, H$_2$O
  features overlap with other features. Optical constants of amorphous
  solid H$_2$O at $T=10$~K \citep{hud93} were used to calculate the
  absorption spectrum of ice spheres \citep{boh83}. Spheres with radii
  of 0.4 \mum\ fit the typical short wavelength profile and peak
  position of the observed 3 \mum\ bands best.  The peak strengths of
  the 6.0 \mum\ bending and 13 \mum\ libration modes relative to the
  3.0 stretching mode of this calculated spectrum are $\sim$7\% weaker
  with respect to the laboratory transmission spectrum \citep{hud93}.
  Also, the libration mode peaks at 12.5 \mum, compared to 13.3
  \mum\ in the laboratory.

\item {\it Silicate feature depth.} The contribution of silicate
  features to the absorption is quantified by the peak optical depth
  of a synthetic silicate spectrum at 9.7~\mum\ ($\tau _{9.7}$). The
  silicate spectrum was calculated for grains small compared to the
  wavelength (Rayleigh limit) using optical constants \citep{dor95} of
  amorphous pyroxene (${\rm Mg}_{0.5}{\rm Fe}_{0.5}{\rm SiO}_3$) and
  olivine (${\rm Mg}_1{\rm Fe}_1{\rm SiO}_4$). The olivine and
  pyroxene spectra were added such that the shape of the silicate
  feature in 2MASS~J21240517+4959100 is matched, i.e.,
  $\tau_{9.7}$(pyroxene)/$\tau_{9.7}$(olivine)=0.62. Note that this
  silicate spectrum is broader than the diffuse medium spectrum
  \citep{kem04}, for which one would need a ratio of 0.20.


\end{itemize}

\begin{figure*}
\includegraphics[angle=90, scale=0.58]{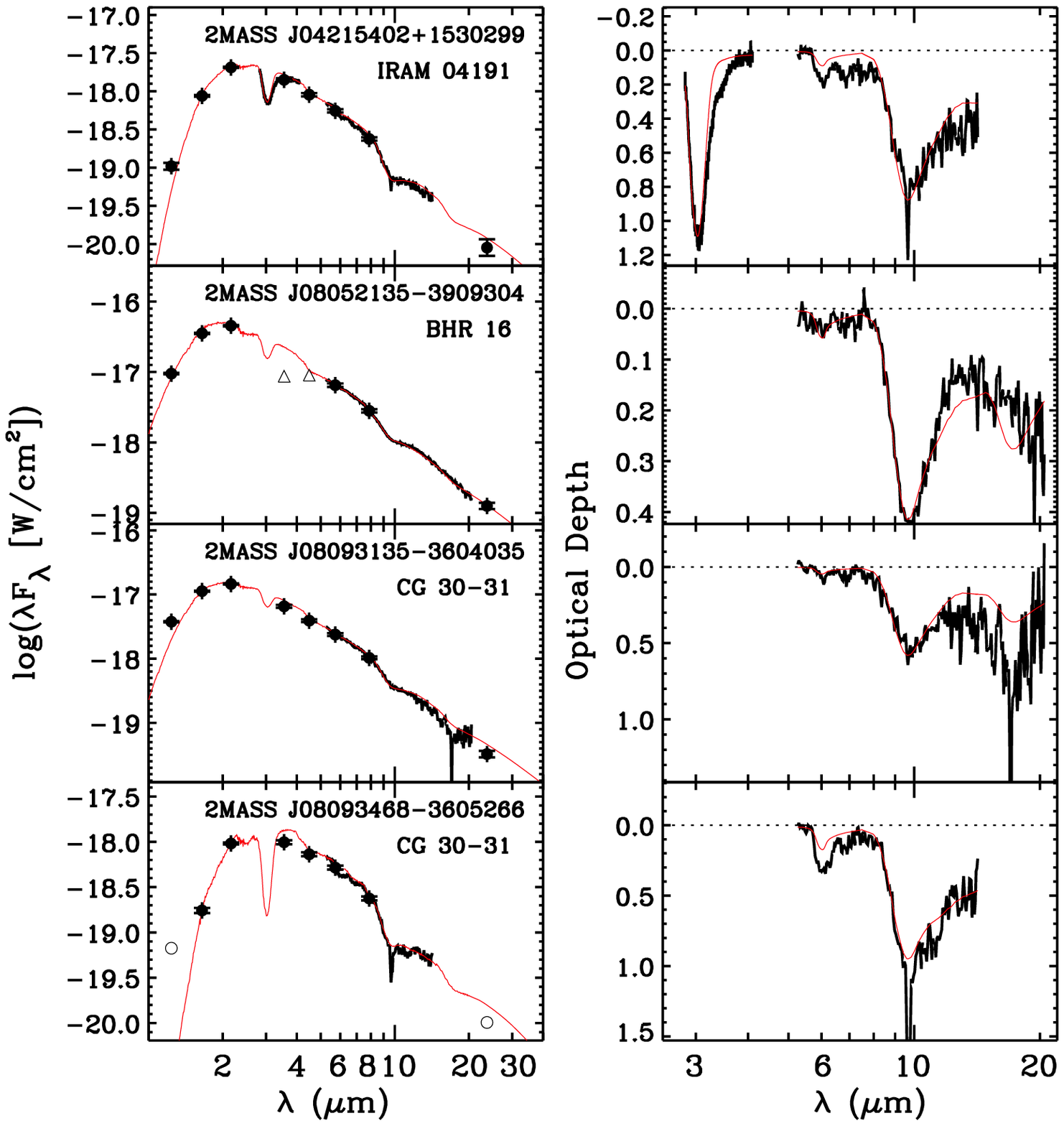}
\includegraphics[angle=90, scale=0.58]{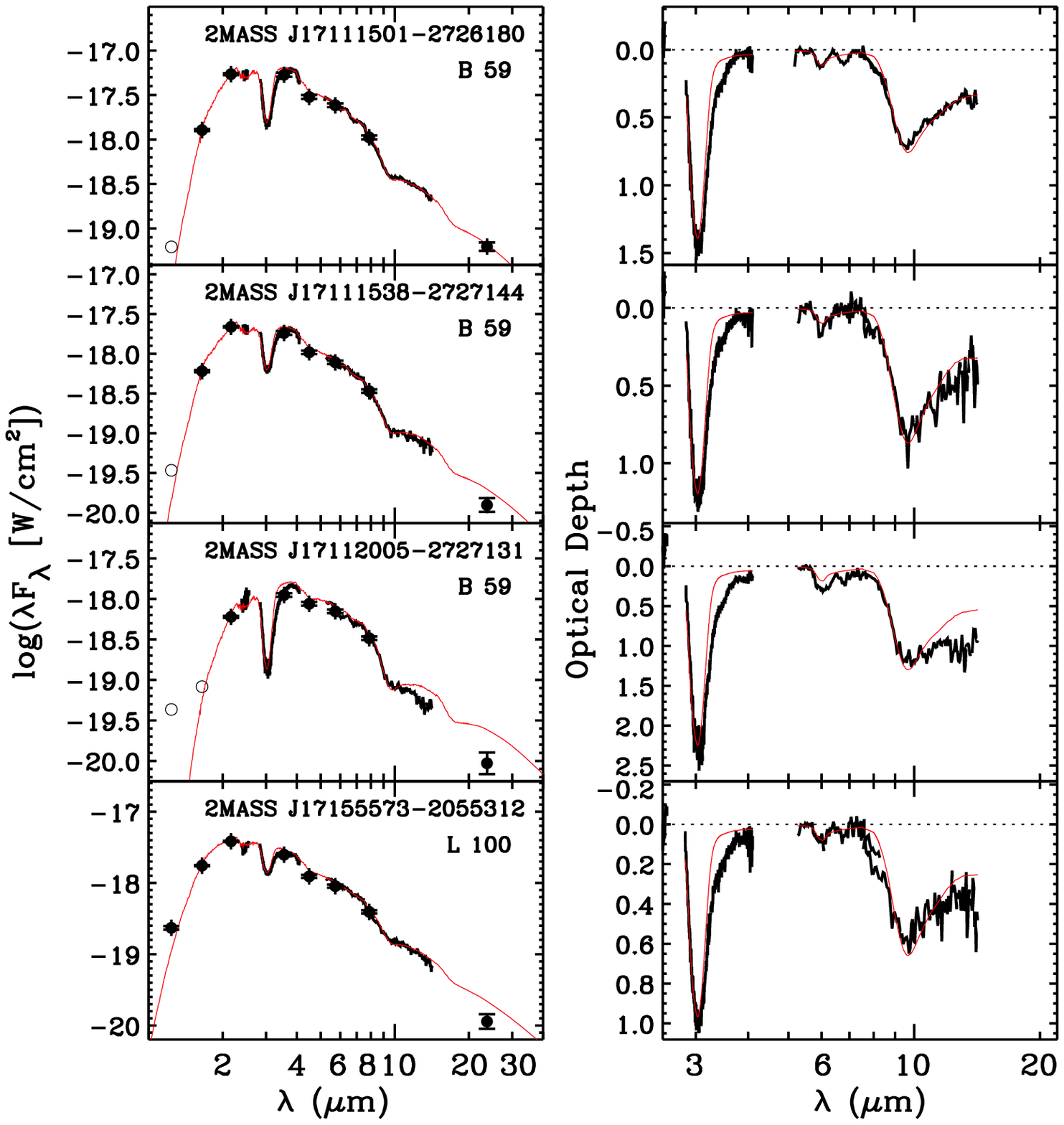}
\includegraphics[angle=90, scale=0.58]{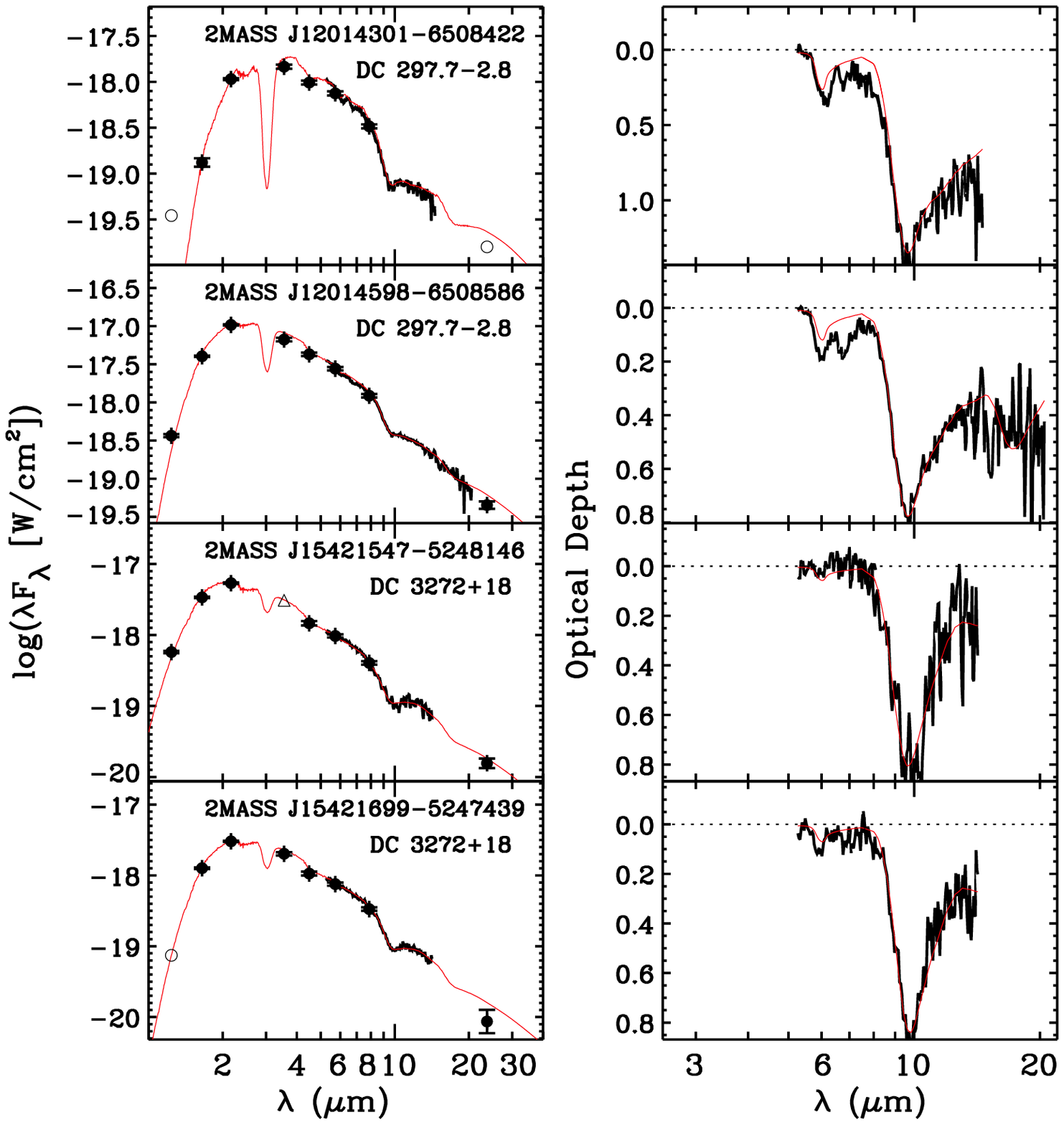}
\includegraphics[angle=90, scale=0.58]{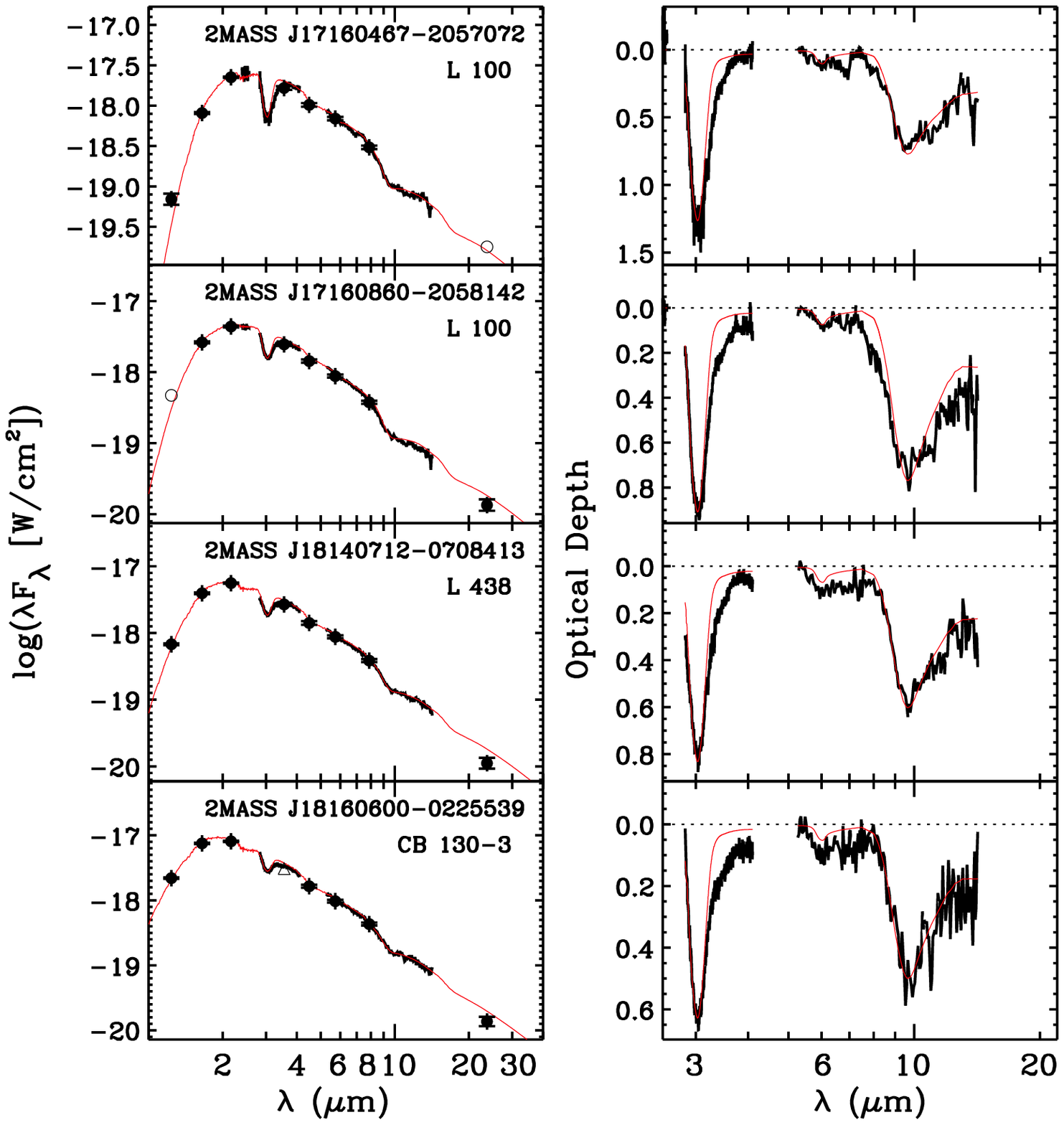}
\caption{{\bf Left panels:} Observed ground-based and {\it
    Spitzer}/IRS spectra combined with broad band photometry (filled
  circles), and lower limits (open triangles) and 3$\sigma$ upper
  limits (open circles) thereof. The red line represents the fitted
  model (\S\ref{sec:cont}). The sources are sorted in increasing Right
  Ascension. The name of the background stars and the cores they trace
  are indicated. {\bf Right panels:} Optical depth spectra, derived
  using the continuum models of the left panels, excluding the modeled
  ice and silicate features. Note the large variation of the relative
  depths of the 3.0 and 9.7~\mum\ bands. The red line indicates the
  modeled H$_2$O ice and silicates spectrum. For clarity, error bars
  of the spectral data points are not shown.}~\label{f:obs1}
\end{figure*}

\setcounter{figure}{0}

\begin{figure*}
\includegraphics[angle=90, scale=0.58]{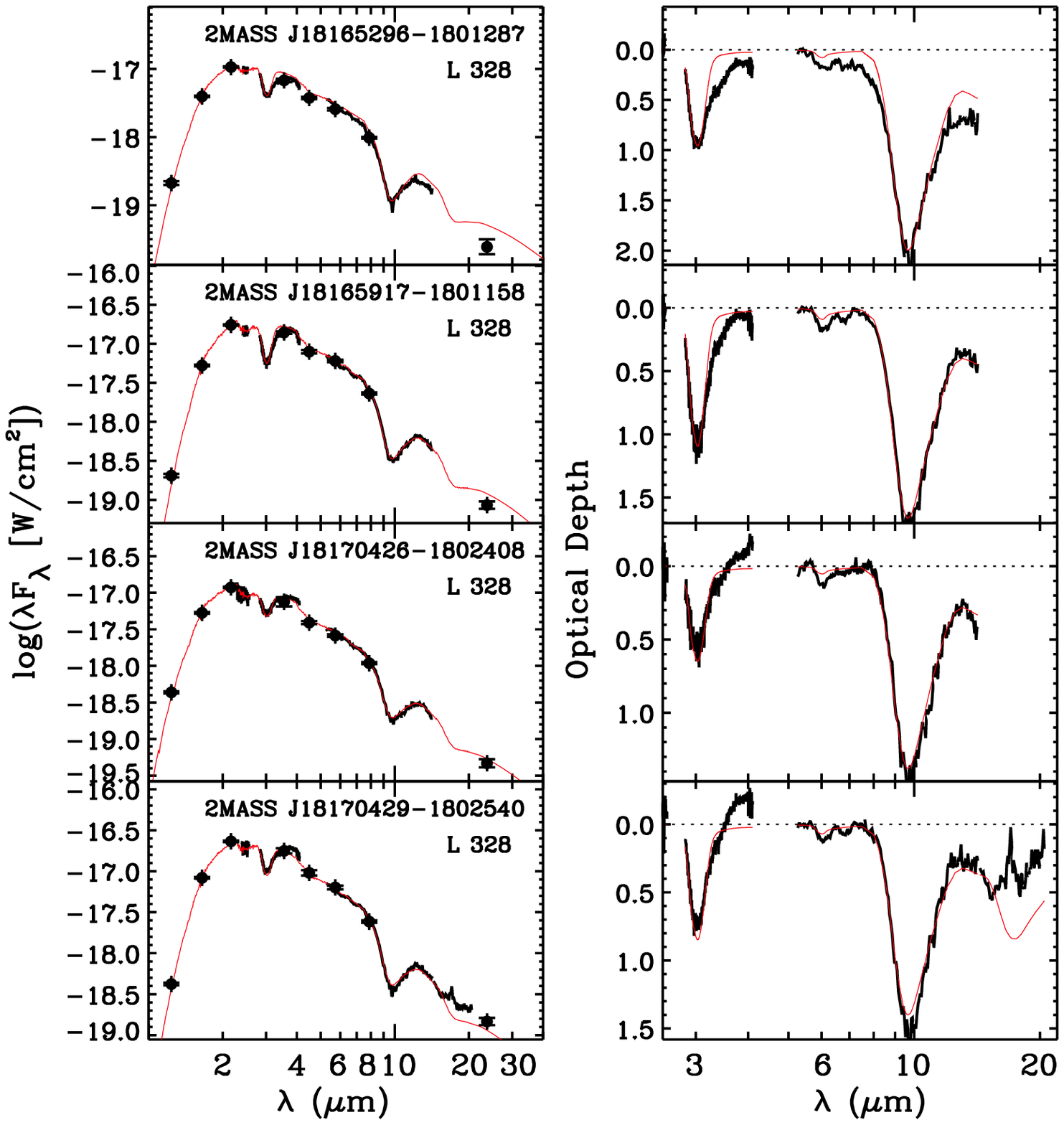}
\includegraphics[angle=90, scale=0.58]{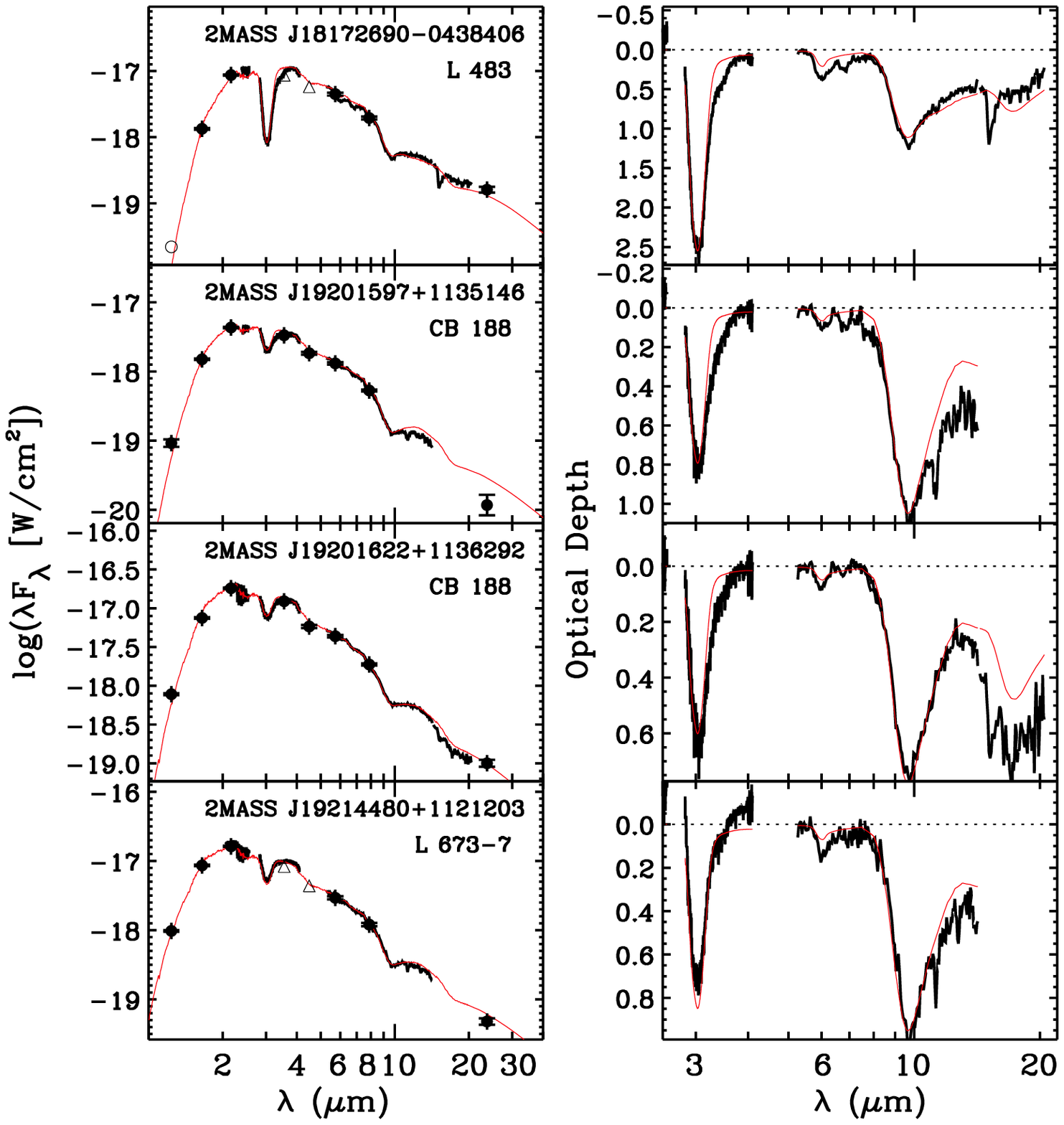}
\includegraphics[angle=90, scale=0.58]{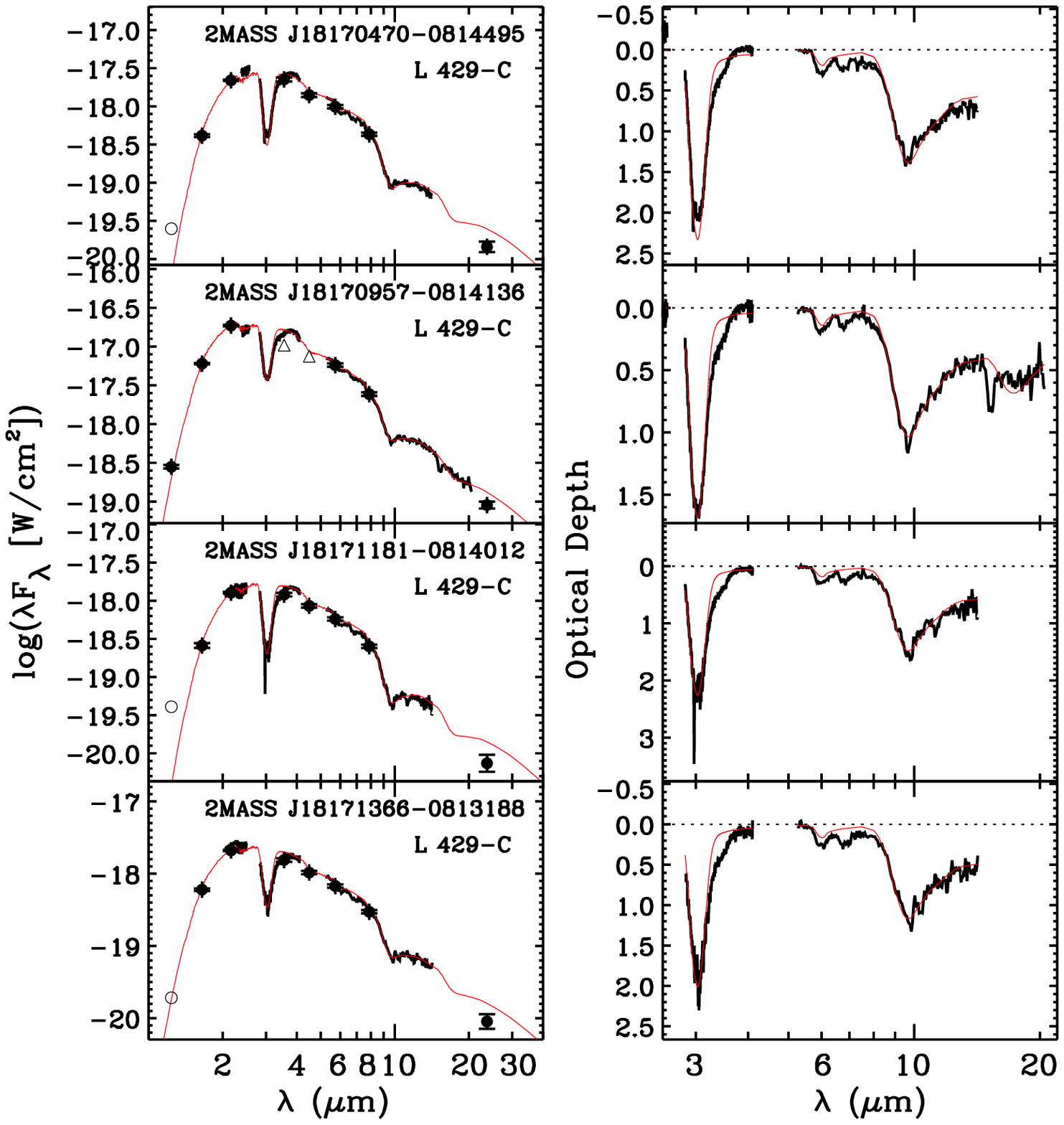}
\includegraphics[angle=90, scale=0.58]{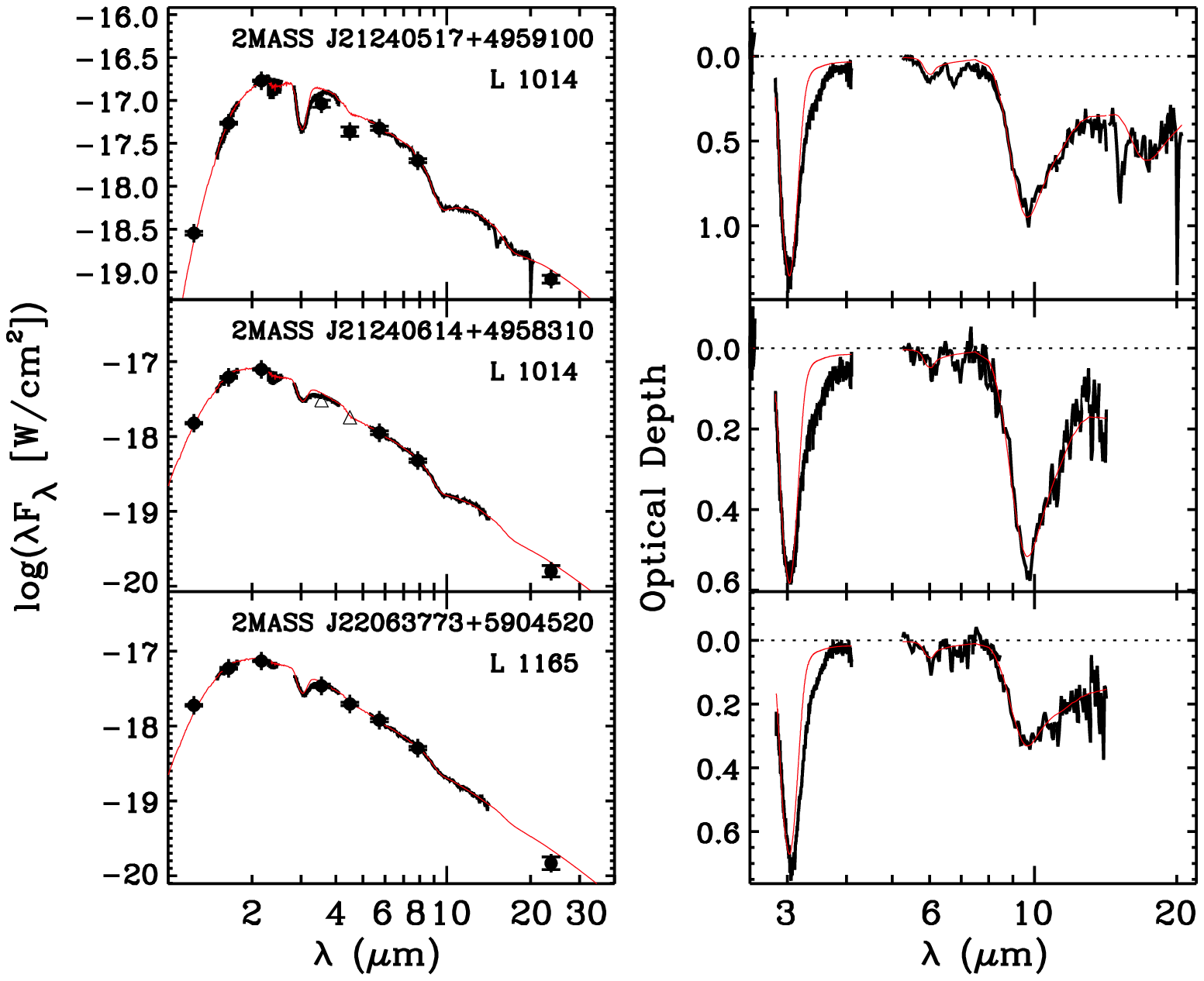}
\caption{(Continuation)}~\label{f:obs3}
\end{figure*}

\begin{deluxetable}{lcc}
\tabletypesize{\scriptsize}
\tablecolumns{3}
\tablewidth{0pc}
\tablecaption{Model Spectra~\label{t:spt}}
\tablehead{
\colhead{Spectral Type}& $T_{\rm eff}$ & \colhead{log({\it g})} \\
\colhead{      }& \colhead{ K      } & \colhead{             } \\}
\startdata
G8 III & 5000 & 2.50\\
K0 III & 4500 & 2.50\\
K3 III & 4500 & 2.00\\
K4 III & 4250 & 2.00\\
K5 III & 4000 & 2.00\\
K7 III & 4000 & 1.50\\
M0 III & 3700 & 1.00\\
M1 III & 3500 & 1.00\\
M3 III & 3200 & 1.00\\
M6 III & 3200 & 0.50\\
M7 III & 3000 & 0.50\\
M8 III & 2800 & 0.50\\
M9 III & 2500 & 0.50\\
\enddata
\tablecomments{Spectral types were assigned using polynomial fits to
  empirical $T_{\rm eff}$ and log({\it g}) data following the method
  of \citet{kan03}.}
\end{deluxetable}

$\chi ^2$ values were derived for all available observables, including
photometric points, ground-based spectra and {\it Spitzer} spectra,
except for those for which it is a priori known that insufficient
information is available: the 3.1-3.7~\mum\ spectral region
(long-wavelength wing 3.0~\mum\ band), the 5.3-7.1~\mum\ region (6.0
and 6.85~\mum\ ice bands), and the IRAC2 photometric point
(4.67~\mum\ CO ice band). The MIPS1 photometric point was not fitted
either, to avoid too much weight to the longer wavelengths, which are
not important for the ice bands studied here (although, in practice,
the final models are in quite good agreement with the MIPS1
photometry; Fig.~\ref{f:obs1}). Because there are many more
spectroscopic than photometric data points, the latter would have much
larger $\chi ^2$ values. Thus the spectroscopic $\chi ^2$ values were
divided by the number of data points and then the ``reduced $\chi ^2$
values'' of the different observables were averaged to represent the
goodness of fit over all observables. In addition, a reduced $\chi ^2$
value was derived for the most crucial observables for spectral type
determination: near-infrared photometry, the shape of near-infrared
spectra and the 4.0-4.1~\mum\ spectral region. The reduced $\chi ^2$
values and parameters of the best fits are reported in
Table~\ref{t:fits} and the fits are plotted in Fig.~\ref{f:obs1} (red
lines in the left panels). Good fits are often obtained. In seven
cases, the reduced $\chi ^2$ values exceed 3 and the main cause for
that is indicated in the notes of Table~\ref{t:fits}. Typically, the
relative scalings of the K-band photometry, and the (partial) K-band
and L-band spectra deviate from the model, but the fits are good at
longer wavelengths.

In seven other cases, no spectra below 5 \mum, and thus no independent
measures of spectral type and $N$(H$_2$O) were available
(Table~\ref{t:sample}). This most strongly affects the $N$(H$_2$O)
determination, as the depth of the 13 \mum\ libration mode is
degenerate with the slope of the applied extinction curve and the
assumed silicate band profiles. An uncertainty in the spectral type
affects the shape and depth of the 5-8 \mum\ absorption features
because of photospheric features that emerge for spectral types later
than M1 III.  In practice, the 8 \mum\ photospheric SiO bands can be
used to constrain the spectral type sufficiently well, however.

\begin{deluxetable*}{lcccccc}
\tabletypesize{\scriptsize}
\tablecolumns{7}
\tablewidth{0pc}
\tablecaption{Continuum Fit Parameters~\label{t:fits}}
\tablehead{
\colhead{Source}& \colhead{Spectral Type\tablenotemark{a}}     &\colhead{$A_{\rm K   }  $\tablenotemark{b}}&
                  \colhead{$\tau_{3.0}$\tablenotemark{c}}         &\colhead{$\tau_{\rm 9.7}$\tablenotemark{d}}&
                  \colhead{$\chi_{\nu}^2$(total)\tablenotemark{e}}&\colhead{$\chi_{\nu}^2$(SpT)\tablenotemark{f}}\\
\colhead{2MASS~J}&\colhead{               } & \colhead{mag          } &  
                  \colhead{               } & \colhead{               } & 
                  \colhead{               } & \colhead{               }\\}
\startdata
$04215402+1530299$ & K0 (G8-K5)           & 3.04       (0.09      )  & 1.09 (0.05)  & 0.88 (0.04     )  & 1.20 & 1.93 \\                                   
$08052135-3909304$ & M0 (K5-M1)$^{\rm g}$ & 1.47       (0.04      )  & $<$0.70$^{\rm h}$ & 0.41 (0.02      )  & 0.29 & \nodata$^{\rm j}$ \\                   
$08093135-3604035$ & K0 (G8-M1)$^{\rm g}$ & 1.89       (0.06      )  & 0.55$^{\rm h}$ (0.25)  & 0.58 (0.02      )  & 0.42 & \nodata$^{\rm j}$ \\              
$08093468-3605266$ & M3 (M0-M7)$^{\rm g}$ & 4.52       (0.14      )  & 2.10$^{\rm h}$ (0.42)  & 0.95 (0.04     )  & 0.58 & \nodata$^{\rm j}$ \\               
$12014301-6508422$ & M3 (M3-M7)$^{\rm g}$ & 5.17       (0.15      )  & 3.20$^{\rm h}$ (0.80)  & 1.35 (0.05     )  & 1.94 & \nodata$^{\rm j}$ \\               
$12014598-6508586$ & K3 (G8-K7)$^{\rm g}$ & 3.07       (0.09      )  & 1.45$^{\rm h}$ (0.25)  & 0.78 (0.03      )  & 0.37 & \nodata$^{\rm j}$ \\              
$15421547-5248146$ & M0 (K7-M1)$^{\rm g}$ & 1.94       (0.06      )  & $<$0.70$^{\rm h}$ & 0.81 (0.03     )  & 0.51 & \nodata$^{\rm j}$ \\                    
$15421699-5247439$ & K7 (K5-M3)$^{\rm g}$ & 2.87       (0.09      )  & 0.85$^{\rm h}$ (0.25)  & 0.84$^{\rm i}$ (0.03     )  & 0.26 & \nodata$^{\rm j}$ \\     
$17111501-2726180$ & M6 (M3-M6)           & 3.91       (0.12      )  & 1.39 (0.07)  & 0.76 (0.03     )  & 1.21 & 1.92 \\                                   
$17111538-2727144$ & M3 (M1-M6)           & 3.53       (0.11      )  & 1.20 (0.06)  & 0.87 (0.03     )  & 1.96 & 1.69 \\                                   
$17112005-2727131$ & M3 (M1-M3)           & 6.00       (0.18      )  & 2.25 (0.11)  & 1.30 (0.05     )  & 4.91$^{\rm k}$ & 5.22 \\                         
$17155573-2055312$ & M1 (M0-M1)           & 2.65       (0.08      )  & 0.97 (0.05)  & 0.66 (0.03     )  & 1.57 & 1.94 \\                                   
$17160467-2057072$ & K7 (K7-M0)           & 3.24       (0.10      )  & 1.27 (0.06)  & 0.77 (0.03     )  & 9.38$^{\rm k}$ & 3.42 \\                         
$17160860-2058142$ & G8 (G8-K4)           & 2.45       (0.07      )  & 0.91 (0.05)  & 0.77 (0.08     )  & 0.72 & 1.28 \\                                   
$18140712-0708413$ & K7 (K5-M0)           & 1.89       (0.06      )  & 0.83 (0.04)  & 0.60 (0.02     )  & 0.64 & 0.72 \\                                   
$18160600-0225539$ & M0 (M0-M1)           & 1.25       (0.04      )  & 0.63 (0.09)  & 0.50 (0.02     )  & 1.77 & 1.76 \\                                   
$18165296-1801287$ & M1                   & 3.00       (0.18      )  & 0.96 (0.10)  & 2.00 (0.12     )  & 1.21 & 1.34 \\                                   
$18165917-1801158$ & M6 (M3-M6)           & 3.34       (0.10      )  & 1.10 (0.05)  & 1.66 (0.07     )  & 0.76 & 1.10 \\                                   
$18170426-1802408$ & M6 (M3-M6)           & 2.50       (0.08      )  & 0.65 (0.03)  & 1.38 (0.06     )  & 1.89 & 4.22 \\                                   
$18170429-1802540$ & M1 (M0-M1)           & 2.96       (0.09      )  & 0.85 (0.09)  & 1.40 (0.06      )  & 2.07 & 1.92 \\                                  
$18170470-0814495$ & M0 (M0-M1)           & 4.27       (0.13      )  & 2.33 (0.12)  & 1.39 (0.06     )  & 12.28$^{\rm k}$ & 2.49 \\                        
$18170957-0814136$ & M1 (M1-M6)           & 3.27       (0.10      )  & 1.69 (0.08)  & 1.04 (0.04      )  & 2.02 & 2.49 \\                                  
$18171181-0814012$ & K7 (K5-M0)           & 4.31       (0.13      )  & 2.26 (0.11)  & 1.49 (0.06     )  & 2.00 & 1.11 \\                                   
$18171366-0813188$ & K7 (K5-M0)           & 3.58       (0.11      )  & 2.02 (0.10)  & 1.17 (0.05     )  & 16.21$^{\rm k}$ & 1.93 \\                        
$18172690-0438406$ & M3 (M1-M6)           & 4.60       (0.14      )  & 2.55 (0.13)  & 1.11 (0.04      )  & 4.53$^{\rm k}$ & 0.78 \\                        
$19201597+1135146$ & M1 (M0-M1)           & 3.14       (0.09      )  & 0.79 (0.04)  & 1.05 (0.04     )  & 1.01 & 0.69 \\                                   
$19201622+1136292$ & M6 (M3-M6)           & 2.50       (0.13      )  & 0.60 (0.03)  & 0.78 (0.03      )  & 0.99 & 1.11 \\                                  
$19214480+1121203$ & M6 (M3-M6)           & 2.10       (0.06      )  & 0.85 (0.04)  & 0.95 (0.04     )  & 7.31$^{\rm k}$ & 1.74 \\                         
$21240517+4959100$ & M1 (M0-M1)           & 3.10       (0.09      )  & 1.30 (0.06)  & 0.95 (0.04      )  & 1.03 & 0.82 \\                                  
$21240614+4958310$ & K4 (K4-K5)           & 1.60       (0.05      )  & 0.58 (0.03)  & 0.52 (0.02     )  & 1.35 & 1.33 \\                                   
$22063773+5904520$ & G8                   & 1.74       (0.09      )  & 0.67 (0.03)  & 0.33 (0.01     )  & 2.95 & 2.48 \\                                   
                   &                      &  &  &  &  &  \\                                                                                                
$04393886+2611266$ & K7 (K5-M1)           & 3.00       (0.09      )  & 1.42 (0.07)  & 0.84 (0.03      )  & 0.80 & 2.61 \\                                  
$18300061+0115201$ & M3 (M3-M6)           & 6.38       (0.19      )  & 1.81 (0.09)  & 1.71 (0.07      )  & 7.30$^{\rm k}$ & 18.79 \\                       
\enddata
\tablenotetext{a}{ Best fitting spectral type that is used throughout
  this work. The uncertainty range is given in parentheses, unless the
  spectral type is more accurate than the closest available
  sub-types. Note that the spectral types are limited to the ones
  listed in Table~\ref{t:spt}. All listed spectral types have
  luminosity class III.}
\tablenotetext{b}{ Extinction in the $K$-band}
\tablenotetext{c}{ Peak absorption optical depth of the 3.0~\mum\ H$_2$O ice band.}
\tablenotetext{d}{ Peak absorption optical depth of the 9.7~\mum\ band of silicates.}
\tablenotetext{e}{ Reduced $\chi^2$ values of the model spectrum fitted to all available photometry and spectra.}
\tablenotetext{f}{ Reduced $\chi^2$ values of the model spectrum to all available near-infrared photometry and spectra ($J$, $H$, $K$, and $L$-band), excluding the 3.0~\mum\ ice band.}
\tablenotetext{g}{ Spectral type determined from wavelengths above 5
  \mum\ only because no near-infrared spectra available.}
\tablenotetext{h}{ $\tau_{3.0}$ uncertain because no near-infrared spectra available.}
\tablenotetext{i}{ 9.7~\mum\ band narrower than for most other sources in the sample.}
\tablenotetext{j}{ No near-infrared spectra available.}
\tablenotetext{k}{ The main reason for large values of
  $\chi_{\nu}^2$(total) is different in individual cases: 2MASS
  J17112005-2727131: only partial K-band spectrum available that does
  not match K-band photometry, and only upper limits to 2MASS J and
  H-band photometry available.  2MASS J17160467-2057072: only partial
  K-band spectrum available that does not match K-band photometry,
  possibly due to flux calibration error of the spectrum.  2MASS
  J18170470-0814495: combined partial K and L-band spectra too high
  compared to model, uncertainty in IRAC1 photometry underestimated?
  2MASS J18171366-0813188: K-band spectrum does not match K-band
  photometry, possibly relative K versus L-band calibration error.
  2MASS J18172690-0438406: model too shallow between the H and
  K-bands, and too steep above 10~\mum. 2MASS J19214480+1121203:
  modeled K- and L-band spectra systematically too low. 2MASS
  J18300061+0115201: no IRAC photometry available for scaling the
  spectra, H and K-band slope steeper than can be modeled.}
\end{deluxetable*}

\begin{figure}
\includegraphics[angle=90, scale=0.55]{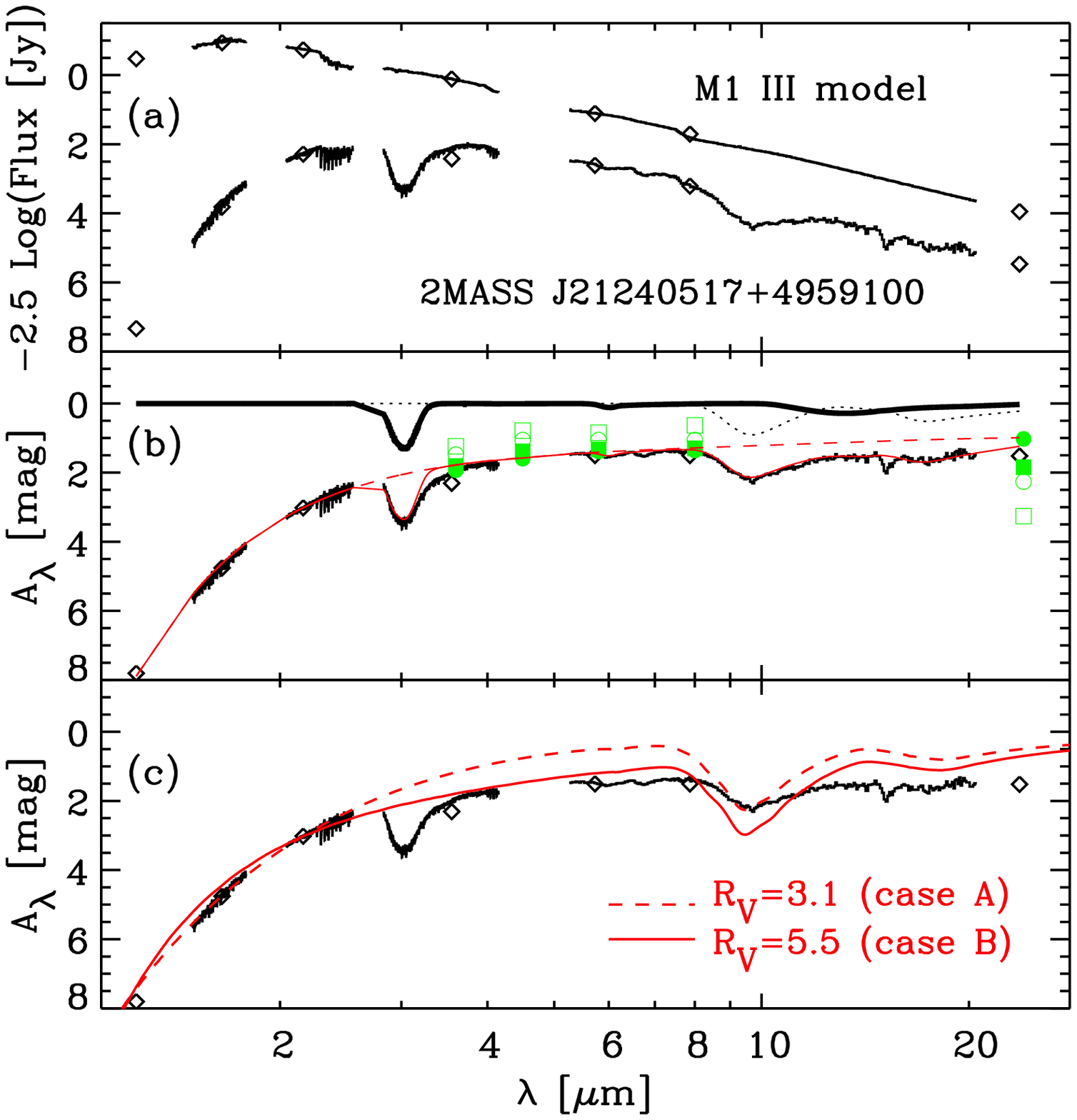}
\caption{Derivation of an extinction curve for the star
  2MASS~J21240517+4959100 behind L1014. {\bf Panel (a)} shows the
  observed spectrum, broad-band photometry (diamonds) and an M1~III
  model spectrum. After subtracting these spectra on a Log scale, the
  resulting extinction curve is shown in {\bf panel (b)}. The solid
  red line is the result of adding spectra of solid H$_2$O (thick
  solid black line) and silicate features (dotted black line) to a
  polynomial fit. The dashed line then represents the feature-free
  extinction curve used in the background star continuum fits. Green
  symbols represent the average extinction in broad band filters
  derived from nearby cloud observations (\citealt{cha09}) for lines
  of sight with $A_{\rm K}<0.5$ (open squares), $0.5<A_{\rm K}<1.0$
  (open circles), $1.0<A_{\rm K}<2.0$ (filled squares), and
  $1.0<A_{\rm K}<2.0$ (filled circles). {\bf Panel (c)} compares the
  empirical extinction curve with the calculated curves of $R_{\rm
    V}$=3.1 (case A) and 5.5 (case B) from
  \citet{wei01}.}~\label{f:ext}
\end{figure}

\subsection{The ``High Resolution'' Extinction Curve}~\label{sec:ext}

\begin{deluxetable}{cc}
\tablecolumns{2}
\tablewidth{0pc}
\tablecaption{Polynomial Coefficients of Feature-Free Extinction
  Curve (Eq.~\ref{eq:pol})~\label{t:pol}}
\tablehead{
\colhead{Coefficient}& Value \\}
\startdata
 a$_0$ &        0.5924\\
 a$_1$ &       -1.8235\\
 a$_2$ &       -1.3020\\
 a$_3$ &        5.9936\\
 a$_4$ &       -5.3429\\
 a$_5$ &        1.2619\\
 a$_6$ &        0.2738\\
 a$_7$ &        0.0069\\
 a$_8$ &       -0.0554\\
\enddata
\end{deluxetable}

The background star spectra can be used to derive a high resolution
extinction curve, in which the ice and dust features can be separated
from the continuum extinction and in which large interpolations
usually applied between, e.g., the IRAC 8~\mum\ and MIPS
24~\mum\ photometric data points are avoided. 2MASS~J21240517+4959100
behind the core L1014 was chosen for this purpose, because high
quality near-infrared spectra are available, indicating an accurate
spectral type of M1~III. The extinction curve was derived as follows:

\begin{equation}
   A_\lambda {\rm [mag]}=-2.5{\rm Log}(F_\lambda{\rm
     (obs)}/F_\lambda{\rm (model)}).
\end{equation}

\noindent The absolute scaling of $F_\lambda{\rm (model)}$ is unknown,
because there are no measurements of the star's distance. To
circumvent this problem, the model was scaled with respect to the
observed spectrum such that the derived $A_\lambda$ follows known
near-infrared extinction curves (i.e.,
$A_\lambda\propto\lambda^{-1.8}$; e.g., \citealt{ind05}).  This
provides the ``high-resolution'' extinction curve shown in
Fig.~\ref{f:ext}b. This curve is remarkably flat. At an $A_{\rm K}$ of
3.02 mag, the extinction is still 1.5 mag at 25~\mum. This confirms
the results of \citet{cha09}, as indicated by the symbols in
Fig.~\ref{f:ext}b, and \citet{mcc09}.  Subsequently, continuum and
feature extinction were separated by adding a laboratory spectrum of
solid H$_2$O (at $T=10$ K; \citealt{hud93}) and a silicates model
spectrum (\S\ref{sec:cont}) to a polynomial fit to feature-free
regions of the form

\begin{equation}
  {\rm Log}(A_\lambda/A_{\rm K})=a_0+a_1{\rm Log(}\lambda)+a_2{\rm
    [{\rm Log}(}\lambda)]^2+...  ~\label{eq:pol}
\end{equation}

with $\lambda$ in \mum\ and $A_\lambda$ in magnitude.  The polynomial
coefficients are listed in Table~\ref{t:pol}, and the curve is plotted
with a dashed red line in Fig.~\ref{f:ext}b.  This feature-free
extinction curve works well for most sources. Examples of exceptions
might be 2MASS J19201597+1135146 (model too shallow $>10$~\mum) and
2MASS J18172690-0438406 (model too steep $>10$~\mum), but no
systematic trend with $A_{\rm K}$ or core environment could be
found. In these cases the silicate model used in the fits may be
responsible for the deviations as well.

\begin{turnpage}
\begin{deluxetable*}{lccccccccccc}
\tabletypesize{\footnotesize}
\setlength{\tabcolsep}{0.00in} 
\tablecolumns{12}
\tablewidth{0pc}
\tablecaption{Optical Depths 5-8~\mum\ Features {\it [publish electronically]}~\label{t:tau}}
\tablehead{
\colhead{Source}& \multicolumn{4}{c}{$\tau_{\rm int}$ [cm$^{-1}$]}   &
                  \colhead{$\tau_{\rm 6.0}$\tablenotemark{e}}      &\colhead{$\tau_{\rm 6.85}$\tablenotemark{f}} &
                  \colhead{$\tau_{\rm C1}$\tablenotemark{g}}      &\colhead{$\tau_{\rm C2}$\tablenotemark{h}} &
                  \colhead{$\tau_{\rm C3}$\tablenotemark{i}}      & \colhead{$\tau_{\rm C4}$\tablenotemark{j}}&
                  \colhead{$\tau_{\rm C5}$\tablenotemark{k}}\\
\colhead{      }&\colhead{5.2-6.4~\mum\tablenotemark{a}}&\colhead{5.2-6.4~\mum\tablenotemark{b}}&
                  \colhead{6.4-7.2~\mum\tablenotemark{c}}&\colhead{6.4-7.2~\mum\tablenotemark{d}}&
                  \colhead{ }      &\colhead{ } &
                  \colhead{ }      &\colhead{ } &
                  \colhead{ }      & \colhead{ }&
                  \colhead{ }\\
\colhead{2MASS~J}&\colhead{               } & \colhead{ minus H$_2$O  } &  
                  \colhead{               } & \colhead{ minus H$_2$O  } & 
                  \colhead{               } & \colhead{               } & 
                  \colhead{               } & \colhead{               } & 
                  \colhead{               } & \colhead{               } & 
                  \colhead{               } \\}
\startdata
$04215402+1530299$  &  26.27   (1.96) &  14.06   (1.96) &  17.48   (2.69) &  12.44   (2.69) &   0.19   (0.04) &   0.15   (0.07) &   0.08   (0.01) &   0.09   (0.01) &   0.10   (0.03) &   0.08   (0.03) &   0.00   (0.05) \\
$08052135-3909304$  &   6.57   (0.94) &  -1.26   (0.94) &   7.54   (1.10) &   4.31   (1.10) &   0.05   (0.02) &   0.06   (0.03) &   0.00   (0.01) &   0.00   (0.01) &   0.03   (0.01) &   0.03   (0.01) &   0.00   (0.02) \\
$08093135-3604035$  &  10.44   (1.02) &   4.30   (1.02) &   9.36   (2.68) &   6.83   (2.68) &   0.10   (0.02) &   0.09   (0.03) &   0.01   (0.01) &   0.04   (0.01) &   0.05   (0.01) &   0.04   (0.01) &   0.00   (0.03) \\
$08093468-3605266$  &  50.63   (1.71) &  27.16   (1.71) &  22.61   (2.24) &  12.94   (2.24) &   0.33   (0.04) &   0.17   (0.05) &   0.14   (0.01) &   0.18   (0.01) &   0.09   (0.02) &   0.08   (0.02) &   0.00   (0.04) \\
$12014301-6508422$  &  49.01   (1.05) &  13.22   (1.05) &  19.23   (1.32) &   4.49   (1.32) &   0.33   (0.02) &   0.19   (0.04) &   0.02   (0.01) &   0.12   (0.01) &   0.10   (0.02) &   0.00   (0.01) &   0.00   (0.04) \\
$12014598-6508586$  &  26.41   (1.09) &  10.20   (1.09) &  20.76   (1.38) &  14.08   (1.38) &   0.18   (0.02) &   0.16   (0.04) &   0.07   (0.01) &   0.06   (0.01) &   0.13   (0.02) &   0.07   (0.01) &   0.00   (0.03) \\
$15421547-5248146$  &   8.36   (2.18) &   0.54   (2.18) &   6.47   (2.54) &   3.25   (2.54) &   0.06   (0.04) &   0.07   (0.06) &   0.01   (0.01) &   0.00   (0.02) &   0.03   (0.02) &   0.00   (0.03) &   0.00   (0.05) \\
$15421699-5247439$  &  15.41   (1.26) &   5.90   (1.26) &  12.40   (1.65) &   8.49   (1.65) &   0.11   (0.03) &   0.10   (0.04) &   0.07   (0.01) &   0.01   (0.01) &   0.06   (0.02) &   0.06   (0.02) &   0.00   (0.03) \\
$17111501-2726180$  &  18.76   (0.64) &   3.17   (0.64) &  13.70   (0.80) &   7.28   (0.80) &   0.14   (0.01) &   0.12   (0.02) &   0.03   (0.01) &   0.02   (0.01) &   0.09   (0.01) &   0.03   (0.01) &   0.00   (0.02) \\
$17111538-2727144$  &  24.14   (2.75) &  10.77   (2.75) &   8.11   (3.48) &   2.60   (3.48) &   0.20   (0.06) &   0.08   (0.09) &   0.06   (0.02) &   0.05   (0.02) &   0.01   (0.04) &   0.03   (0.04) &   0.00   (0.07) \\
$17112005-2727131$  &  46.71   (1.95) &  21.56   (1.95) &  20.81   (2.21) &  10.44   (2.21) &   0.30   (0.04) &   0.20   (0.06) &   0.11   (0.01) &   0.13   (0.01) &   0.13   (0.02) &   0.04   (0.02) &   0.00   (0.04) \\
$17155573-2055312$  &  14.83   (1.60) &   4.00   (1.60) &   8.26   (1.97) &   3.80   (1.97) &   0.11   (0.03) &   0.08   (0.05) &   0.04   (0.01) &   0.02   (0.01) &   0.06   (0.02) &   0.00   (0.02) &   0.00   (0.04) \\
$17160467-2057072$  &  14.30   (2.48) &   0.15   (2.48) &  19.51   (3.38) &  13.67   (3.38) &   0.10   (0.05) &   0.18   (0.08) &   0.00   (0.02) &   0.01   (0.02) &   0.11   (0.03) &   0.10   (0.03) &   0.00   (0.07) \\
$17160860-2058142$  &  11.18   (1.31) &   1.00   (1.31) &   7.75   (1.78) &   3.55   (1.78) &   0.08   (0.03) &   0.08   (0.05) &   0.00   (0.01) &   0.01   (0.01) &   0.05   (0.02) &   0.01   (0.02) &   0.00   (0.03) \\
$18140712-0708413$  &  13.30   (1.77) &   3.98   (1.77) &  10.78   (1.61) &   6.94   (1.61) &   0.09   (0.03) &   0.07   (0.04) &   0.02   (0.01) &   0.04   (0.01) &   0.04   (0.02) &   0.04   (0.02) &   0.00   (0.04) \\
$18160600-0225539$  &  14.64   (1.66) &   7.60   (1.66) &  11.23   (2.21) &   8.33   (2.21) &   0.10   (0.03) &   0.09   (0.06) &   0.05   (0.01) &   0.04   (0.01) &   0.04   (0.02) &   0.06   (0.02) &   0.00   (0.04) \\
$18165296-1801287$  &  19.54   (0.92) &   8.81   (0.92) &   8.00   (1.24) &   3.58   (1.24) &   0.13   (0.02) &   0.09   (0.03) &   0.04   (0.01) &   0.06   (0.01) &   0.06   (0.01) &   0.01   (0.01) &   0.00   (0.03) \\
$18165917-1801158$  &  24.64   (1.05) &  12.37   (1.05) &  12.63   (1.22) &   7.58   (1.22) &   0.17   (0.02) &   0.11   (0.03) &   0.05   (0.01) &   0.10   (0.01) &   0.08   (0.01) &   0.04   (0.01) &   0.00   (0.02) \\
$18170426-1802408$  &  23.25   (1.18) &  15.98   (1.18) &  14.30   (1.46) &  11.31   (1.46) &   0.17   (0.02) &   0.10   (0.04) &   0.07   (0.01) &   0.12   (0.01) &   0.08   (0.01) &   0.06   (0.02) &   0.00   (0.03) \\
$18170429-1802540$  &  17.86   (2.29) &   8.34   (2.29) &   9.63   (2.10) &   5.72   (2.10) &   0.13   (0.05) &   0.09   (0.05) &   0.03   (0.01) &   0.07   (0.02) &   0.06   (0.02) &   0.04   (0.02) &   0.00   (0.05) \\
$18170470-0814495$  &  40.68   (1.48) &  14.60   (1.48) &  23.81   (1.85) &  13.06   (1.85) &   0.28   (0.03) &   0.21   (0.05) &   0.11   (0.01) &   0.07   (0.01) &   0.15   (0.02) &   0.06   (0.02) &   0.00   (0.04) \\
$18170957-0814136$  &  27.44   (1.10) &   8.56   (1.10) &  18.09   (1.22) &  10.31   (1.22) &   0.19   (0.02) &   0.16   (0.03) &   0.07   (0.01) &   0.03   (0.01) &   0.11   (0.01) &   0.06   (0.01) &   0.00   (0.02) \\
$18171181-0814012$  &  41.27   (1.68) &  15.97   (1.68) &  26.33   (2.23) &  15.90   (2.23) &   0.28   (0.04) &   0.23   (0.06) &   0.12   (0.01) &   0.07   (0.01) &   0.15   (0.02) &   0.11   (0.02) &   0.00   (0.04) \\
$18171366-0813188$  &  34.05   (1.63) &  11.50   (1.63) &  25.80   (2.12) &  16.50   (2.12) &   0.24   (0.04) &   0.21   (0.06) &   0.08   (0.01) &   0.06   (0.01) &   0.15   (0.02) &   0.10   (0.02) &   0.00   (0.04) \\
$18172690-0438406$  &  52.65   (1.15) &  24.10   (1.15) &  28.26   (1.34) &  16.49   (1.34) &   0.34   (0.03) &   0.23   (0.04) &   0.14   (0.01) &   0.15   (0.01) &   0.16   (0.02) &   0.09   (0.01) &   0.00   (0.03) \\
$19201597+1135146$  &  14.91   (1.20) &   6.04   (1.20) &  10.25   (1.72) &   6.59   (1.72) &   0.10   (0.03) &   0.09   (0.04) &   0.03   (0.01) &   0.04   (0.01) &   0.08   (0.02) &   0.04   (0.02) &   0.00   (0.04) \\
$19201622+1136292$  &  11.74   (0.76) &   5.00   (0.76) &   4.40   (0.77) &   1.62   (0.77) &   0.09   (0.01) &   0.05   (0.02) &   0.04   (0.01) &   0.03   (0.01) &   0.03   (0.01) &   0.00   (0.01) &   0.00   (0.02) \\
$19214480+1121203$  &  19.51   (1.72) &  10.01   (1.72) &  10.35   (2.22) &   6.43   (2.22) &   0.15   (0.04) &   0.09   (0.06) &   0.04   (0.01) &   0.08   (0.01) &   0.02   (0.02) &   0.06   (0.02) &   0.00   (0.04) \\
$21240517+4959100$  &  19.44   (1.23) &   4.93   (1.23) &  13.95   (1.31) &   7.97   (1.31) &   0.13   (0.02) &   0.14   (0.03) &   0.03   (0.01) &   0.02   (0.01) &   0.12   (0.01) &   0.02   (0.01) &   0.00   (0.02) \\
$21240614+4958310$  &   8.54   (1.26) &   2.01   (1.26) &   8.12   (1.54) &   5.44   (1.54) &   0.08   (0.03) &   0.08   (0.04) &   0.00   (0.01) &   0.02   (0.01) &   0.03   (0.02) &   0.06   (0.02) &   0.00   (0.03) \\
$22063773+5904520$  &  10.22   (1.31) &   2.70   (1.31) &   9.70   (1.73) &   6.61   (1.73) &   0.09   (0.03) &   0.08   (0.05) &   0.00   (0.01) &   0.03   (0.01) &   0.05   (0.02) &   0.05   (0.02) &   0.00   (0.03) \\
 &  &  &  &  &  &  &  &  &  &  &  \\
$04393886+2611266$  &  20.66   (0.22) &   4.78   (0.22) &  13.48   (0.29) &   6.93   (0.29) &   0.15   (0.01) &   0.12   (0.01) &   0.03   (0.01) &   0.02   (0.01) &   0.08   (0.01) &   0.04   (0.01) &   0.00   (0.01) \\
$18300061+0115201$  &  33.96   (1.25) &  13.77   (1.25) &  20.82   (0.89) &  12.51   (0.89) &   0.26   (0.03) &   0.20   (0.02) &   0.10   (0.01) &   0.08   (0.01) &   0.14   (0.01) &   0.08   (0.01) &   0.00   (0.03) \\
\enddata
\tablecomments{Uncertainties in parentheses based on statistical errors in the spectra only,
 unless noted otherwise below.}
\tablenotetext{a}{ integrated optical depth between 5.2-6.4~\mum\ in wavenumber units}
\tablenotetext{b}{ integrated optical depth between 5.2-6.4~\mum\ in wavenumber units, after subtraction of a laboratory spectrum of pure H$_2$O ice}
\tablenotetext{c}{ integrated optical depth between 6.4-7.2~\mum\ in wavenumber units}
\tablenotetext{d}{ integrated optical depth between 6.4-7.2~\mum\ in wavenumber units, after subtraction of a laboratory spectrum of pure H$_2$O ice}
\tablenotetext{e}{ peak optical depth at 6.0~\mum}
\tablenotetext{f}{ peak optical depth at 6.85~\mum}
\tablenotetext{g}{ peak optical depth component C1 \citep{boo08}}
\tablenotetext{h}{ peak optical depth component C2 }
\tablenotetext{i}{ peak optical depth component C3 }
\tablenotetext{j}{ peak optical depth component C4 }
\tablenotetext{k}{ peak optical depth component C5. Uncertainty includes uncertainty in $N$(H$_2$O) as this affects the baseline level.}
\end{deluxetable*}
\end{turnpage}

\begin{turnpage}
\begin{deluxetable*}{llccrrrrrrrrrr}
\tabletypesize{\footnotesize}
\setlength{\tabcolsep}{0.00in} 
\tablecolumns{14}
\tablewidth{0pc}
\tablecaption{Column Densities of the Ices~\label{t:colden}}
\tablehead{
\colhead{Source}& \colhead{$N$(H$_2$O)\tablenotemark{a}}     & 
                  \multicolumn{2}{c}{$N$(NH$_4^+$)\tablenotemark{b}}   &
                  \multicolumn{2}{c}{$N$(CO$_2$)} &
                  \multicolumn{2}{c}{$N$(CH$_3$OH)\tablenotemark{c}} &
                  \multicolumn{2}{c}{$N$(HCOOH)} &
                  \multicolumn{2}{c}{$N$(CH$_4$)} &
                  \multicolumn{2}{c}{$N$(NH$_3$)} \\
\colhead{2MASS~J}& \colhead{$10^{18}$ } & 
                   \colhead{$10^{17}$ } & 
                   \colhead{      } &
                   \colhead{$10^{17}$ } & 
                   \colhead{      } &
                   \colhead{$10^{17}$ } & 
                   \colhead{      } &
                   \colhead{$10^{17}$ } & 
                   \colhead{      } &
                   \colhead{$10^{17}$ } & 
                   \colhead{      } &
                   \colhead{$10^{17}$ } & 
                   \colhead{      } \\
\colhead{       }& \colhead{ \sqcm} & 
                   \colhead{ \sqcm} & 
                   \colhead{\%H$_2$O      } &
                   \colhead{ \sqcm} & 
                   \colhead{\%H$_2$O      } &
                   \colhead{ \sqcm} & 
                   \colhead{\%H$_2$O      } &
                   \colhead{ \sqcm} & 
                   \colhead{\%H$_2$O      } &
                   \colhead{ \sqcm} & 
                   \colhead{\%H$_2$O      } &
                   \colhead{ \sqcm} & 
                   \colhead{\%H$_2$O      } \\}
\startdata
  $04215402+1530299  $ &       1.84                  (0.20) &     2.73 (0.61) &    14.80 (3.71) &  \nodata        &  \nodata        &  $<$1.55        &  $<$9.46        &  $<$0.98        &  $<$6.03        &  $<$1.54        &  $<$9.45        &  $<$5.46        & $<$33.38        \\
  $08052135-3909304  $ &    $<$1.18       \tablenotemark{d} &     0.98 (0.24) &  $>$8.29        &     3.79        & $>$32.11        &  $<$0.50        &  \nodata        &  $<$0.23        &  \nodata        &  $<$1.02        &  \nodata        &  $<$2.00        &  \nodata        \\
  $08093135-3604035  $ &       0.92 (0.42)\tablenotemark{d} &     1.55 (0.60) &    16.71 (10.1) & $<$11.31        & $<$226.0        &  $<$2.00        & $<$39.97        &  $<$0.23        &  $<$4.61        &  $<$2.15        & $<$43.03        &  $<$3.00        & $<$59.95        \\
  $08093468-3605266  $ &       3.54 (0.79)\tablenotemark{d} &     2.94 (0.50) &     8.29 (2.34) &  \nodata        &  \nodata        & $<$10.00        & $<$36.34        &  $<$1.63        &  $<$5.94        &  $<$1.86        &  $<$6.79        &  $<$4.13        & $<$15.04        \\
  $12014301-6508422  $ &       5.40 (1.45)\tablenotemark{d} &     1.02 (0.30) &     1.89 (0.75) &  \nodata        &  \nodata        &  $<$8.52        & $<$21.59        &  $<$0.29        &  $<$0.75        &  $<$1.15        &  $<$2.92        &  $<$9.00        & $<$22.80        \\
  $12014598-6508586  $ &       2.44 (0.48)\tablenotemark{d} &     3.20 (0.31) &    13.07 (2.87) &     8.34 (1.91) &    34.11 (10.3) &  $<$2.00        & $<$10.18        &  $<$0.81        &  $<$4.17        &  $<$1.01        &  $<$5.17        &  $<$4.00        & $<$20.36        \\
  $15421547-5248146  $ &    $<$1.18       \tablenotemark{d} &     0.73 (0.57) &  $>$6.25        &  \nodata        &  \nodata        &  $<$5.34        &  \nodata        &  $<$0.44        &  \nodata        &  $<$2.03        &  \nodata        &  $<$3.64        &  \nodata        \\
  $15421699-5247439  $ &       1.43 (0.44)\tablenotemark{d} &     1.92 (0.37) &    13.45 (4.88) &  \nodata        &  \nodata        &  $<$2.74        & $<$27.62        &  $<$0.81        &  $<$8.22        &  $<$2.08        & $<$20.99        &  $<$4.06        & $<$40.83        \\
  $17111501-2726180  $ &       2.35                  (0.26) &     1.52 (0.18) &     6.49 (1.06) &  \nodata        &  \nodata        &  $<$1.60        &  $<$7.65        &  $<$0.35        &  $<$1.70        &  $<$0.57        &  $<$2.75        &  $<$1.32        &  $<$6.32        \\
  $17111538-2727144  $ &       2.01                  (0.22) &  $<$2.37        & $<$13.24        &  \nodata        &  \nodata        &  $<$2.00        & $<$11.15        &  $<$0.76        &  $<$4.27        &  $<$2.58        & $<$14.41        &  $<$3.00        & $<$16.73        \\
  $17112005-2727131  $ &       3.79                  (0.42) &     2.17 (0.50) &     5.71 (1.46) &  \nodata        &  \nodata        &  $<$2.00        &  $<$5.92        &  $<$1.34        &  $<$3.99        &  $<$1.39        &  $<$4.14        &  $<$9.00        & $<$26.68        \\
  $17155573-2055312  $ &       1.63                  (0.18) &     0.77 (0.44) &     4.74 (2.79) &  \nodata        &  \nodata        &  $<$0.86        &  $<$5.94        &  $<$0.42        &  $<$2.90        &  $<$1.41        &  $<$9.71        &  $<$4.00        & $<$27.54        \\
  $17160467-2057072  $ &       2.13                  (0.23) &     2.99 (0.76) &    13.99 (3.91) &  \nodata        &  \nodata        &  $<$1.39        &  $<$7.32        &  $<$0.58        &  $<$3.09        &  $<$1.81        &  $<$9.54        &  $<$4.00        & $<$21.06        \\
  $17160860-2058142  $ &       1.53                  (0.17) &     0.72 (0.40) &     4.72 (2.69) &  \nodata        &  \nodata        &  $<$0.43        &  $<$3.17        &  $<$0.31        &  $<$2.32        &  $<$1.10        &  $<$8.10        &  $<$4.00        & $<$29.30        \\
  $18140712-0708413  $ &       1.40                  (0.15) &     1.40 (0.36) &     9.99 (2.83) &  \nodata        &  \nodata        &     0.70 (0.20) &     4.97 (1.52) &  $<$0.30        &  $<$2.46        &  $<$1.05        &  $<$8.45        &  $<$3.50        & $<$27.99        \\
  $18160600-0225539  $ &       1.06                  (0.19) &     1.83 (0.50) &    17.29 (5.65) &  \nodata        &  \nodata        &  $<$0.76        &  $<$8.73        &  $<$0.56        &  $<$6.47        &  $<$1.64        & $<$18.86        &  $<$3.50        & $<$40.18        \\
  $18165296-1801287  $ &       1.62                  (0.22) &     0.72 (0.28) &     4.48 (1.84) &  \nodata        &  \nodata        &  $<$1.36        &  $<$9.82        &  $<$0.48        &  $<$3.49        &  $<$1.20        &  $<$8.62        &  $<$2.48        & $<$17.87        \\
  $18165917-1801158  $ &       1.85                  (0.20) &     1.62 (0.27) &     8.76 (1.78) &  \nodata        &  \nodata        &  $<$0.85        &  $<$5.21        &  $<$0.56        &  $<$3.43        &  $<$0.71        &  $<$4.36        &  $<$3.32        & $<$20.22        \\
  $18170426-1802408  $ &       1.09                  (0.12) &     2.51 (0.33) &    22.90 (3.96) &  \nodata        &  \nodata        &  $<$0.83        &  $<$8.60        &  $<$0.83        &  $<$8.61        &  $<$0.96        &  $<$9.85        &  $<$3.72        & $<$38.23        \\
  $18170429-1802540  $ &       1.43                  (0.20) &     1.22 (0.47) &     8.52 (3.53) &     3.37 (1.95) &    23.55 (14.0) &  $<$0.77        &  $<$6.29        &  $<$0.37        &  $<$3.05        &  $<$0.90        &  $<$7.32        &  $<$3.67        & $<$29.82        \\
  $18170470-0814495  $ &       3.93                  (0.44) &     2.41 (0.42) &     6.13 (1.26) &  \nodata        &  \nodata        &     2.49 (0.60) &     6.34 (1.69) &  $<$1.28        &  $<$3.66        &  $<$1.80        &  $<$5.14        &  $<$7.00        & $<$20.00        \\
  $18170957-0814136  $ &       2.85                  (0.31) &     1.82 (0.27) &     6.39 (1.20) &    12.29 (1.29) &    43.12 (6.63) &     2.69 (0.58) &     9.46 (2.30) &  $<$0.89        &  $<$3.53        &  $<$0.97        &  $<$3.84        &  $<$3.16        & $<$12.48        \\
  $18171181-0814012  $ &       3.81                  (0.42) &     2.80 (0.50) &     7.35 (1.55) &  \nodata        &  \nodata        &     4.42 (0.86) &    11.58 (2.60) &  $<$1.48        &  $<$4.38        &  $<$1.42        &  $<$4.19        &  $<$6.71        & $<$19.80        \\
  $18171366-0813188  $ &       3.40                  (0.38) &     3.09 (0.48) &     9.08 (1.74) &  \nodata        &  \nodata        &     3.48 (0.86) &    10.24 (2.78) &  $<$0.99        &  $<$3.29        &  $<$1.24        &  $<$4.12        &  $<$4.83        & $<$15.97        \\
  $18172690-0438406  $ &       4.31                  (0.48) &     3.02 (0.30) &     7.01 (1.05) &    18.86 (1.72) &    43.75 (6.31) &     3.63 (0.65) &     8.44 (1.79) &  $<$1.63        &  $<$4.28        &  $<$0.95        &  $<$2.48        &  $<$7.00        & $<$18.28        \\
  $19201597+1135146  $ &       1.34                  (0.14) &     1.42 (0.39) &    10.64 (3.14) &  \nodata        &  \nodata        &  $<$0.70        &  $<$5.95        &  $<$0.37        &  $<$3.12        &  $<$1.15        &  $<$9.69        &  $<$6.58        & $<$55.33        \\
  $19201622+1136292  $ &       1.01                  (0.11) &     0.31 (0.17) &     3.09 (1.74) &     3.86 (1.07) &    38.08 (11.3) &  $<$0.70        &  $<$7.77        &  $<$0.50        &  $<$5.58        &  $<$0.58        &  $<$6.49        &  $<$2.00        & $<$22.15        \\
  $19214480+1121203  $ &       1.43                  (0.16) &     1.38 (0.50) &     9.65 (3.68) &  \nodata        &  \nodata        &  $<$1.61        & $<$12.64        &  $<$0.49        &  $<$3.88        &  $<$1.63        & $<$12.85        &  $<$2.57        & $<$20.22        \\
  $21240517+4959100  $ &       2.19                  (0.24) &     1.48 (0.29) &     6.76 (1.55) &     8.24 (2.69) &    37.60 (13.0) &     1.54 (0.38) &     7.04 (1.90) &  $<$0.40        &  $<$2.07        &  $<$1.83        &  $<$9.43        &  $<$4.20        & $<$21.57        \\
  $21240614+4958310  $ &       0.98                  (0.11) &     1.18 (0.34) &    12.01 (3.78) &  \nodata        &  \nodata        &  $<$0.59        &  $<$6.83        &  $<$0.27        &  $<$3.13        &  $<$1.14        & $<$13.04        &  $<$2.50        & $<$28.57        \\
  $22063773+5904520  $ &       1.13                  (0.12) &     1.44 (0.39) &    12.70 (3.74) &  \nodata        &  \nodata        &  $<$0.40        &  $<$3.98        &  $<$0.31        &  $<$3.10        &  $<$1.11        & $<$11.02        &  $<$2.30        & $<$22.82        \\
 &  &  &  &  &  &  &  &  &  &  &  &  \\
  $04393886+2611266  $ &       2.39                  (0.26) &     1.44 (0.06) &     6.03 (0.72) &     7.11 (0.05) &    29.66 (3.32) &  $<$0.25        &  $<$1.17        &  $<$0.38        &  $<$1.81        &  $<$0.82        &  $<$3.84        &  $<$1.45        &  $<$6.85        \\
  $18300061+0115201  $ &       3.04                  (0.34) &     2.68 (0.20) &     8.79 (1.18) &    11.93 (0.21) &    39.14 (4.43) &  $<$2.10        &  $<$7.75        &  $<$1.18        &  $<$4.36        &  $<$0.58        &  $<$2.14        &  $<$6.39        & $<$23.61        \\
\enddata

\tablecomments{Column densities were determined using the intrinsic
  integrated band strengths listed in
  Table~\ref{t:avalues}. Uncertainties (1$\sigma$) are indicated in
  brackets and upper limits are of 3$\sigma$ significance.}

\tablenotetext{a}{An uncertainty of 10\% in the intrinsic integrated
  band strength is taken into account in the listed column density
  uncertainties.}

\tablenotetext{b}{Assuming both C3 and C4 components \citep{boo08} are
  due to NH$_4^+$ (\S\ref{sec:col}). The contribution by the CH$_3$OH
  C-H bending mode has been subtracted for sources with CH$_3$OH
  detections.}

\tablenotetext{c}{The uncertainties in the CH$_3$OH column densities
  do not reflect interference with photospheric absorption lines
  (Fig.~\ref{f:ch3oh}). CH$_3$OH detections toward
  2MASS~J18171181-0814012 and 2MASS~J18171366-0813188 are most
  reliable.}

\tablenotetext{d}{No $L$-band spectrum available for H$_2$O ice column
  density determination. The 13~\mum\ H$_2$O libration mode was used
  instead. These values are uncertain, and the listed uncertainties
  are estimates.}

\end{deluxetable*}
\end{turnpage}

\begin{deluxetable*}{llcl}
\tabletypesize{\scriptsize}
\tablecolumns{4}
\tablewidth{0pc}
\tablecaption{Intrinsic Integrated Band Strengths~\label{t:avalues}}
\tablehead{
\colhead{Molecule}& \colhead{Mode}             & \colhead{$A$}        & \colhead{Reference}\\
\colhead{        }& \colhead{    }             & \colhead{cm/molecule}& \colhead{         } \\}
\startdata
H$_2$O            & 3.0 \mum\  stretch         & $2.0\times 10^{-16}$  & \citet{hag81}\\ 
NH$_4^+$          & 6.8 \mum\  deformation     & $4.4\times 10^{-17}$  & \citet{sch03}\\ 
CO$_2$            & 15.0 \mum\ bend            & $1.1\times 10^{-17}$  & \citet{ger95}\\ 
CH$_3$OH          & 3.53 \mum\ C-H stretch     & $5.6\times 10^{-18}$  & \citet{ker99}\\ 
CH$_3$OH          & 9.7 \mum\  O-H stretch     & $1.6\times 10^{-17}$  & \citet{ker99}\\  
HCOOH             & 5.85 \mum\ C=O stretch     & $6.7\times 10^{-17}$  & \citet{sch99}\\ 
HCOOH             & 7.25 \mum\ C-H deformation & $2.6\times 10^{-18}$  & \citet{sch99}\\ 
CH$_4$            & 7.68 \mum\ deformation     & $7.3\times 10^{-18}$  & \citet{boo97}\\ 
NH$_3$            & 8.9 \mum\  umbrella        & $1.3\times 10^{-17}$  & \citet{ker99}\\ 
\enddata
\end{deluxetable*}

\subsection{Absorption Band Strengths and Column Densities}~\label{sec:col}

\begin{figure}[b]
\includegraphics[angle=90, scale=0.55]{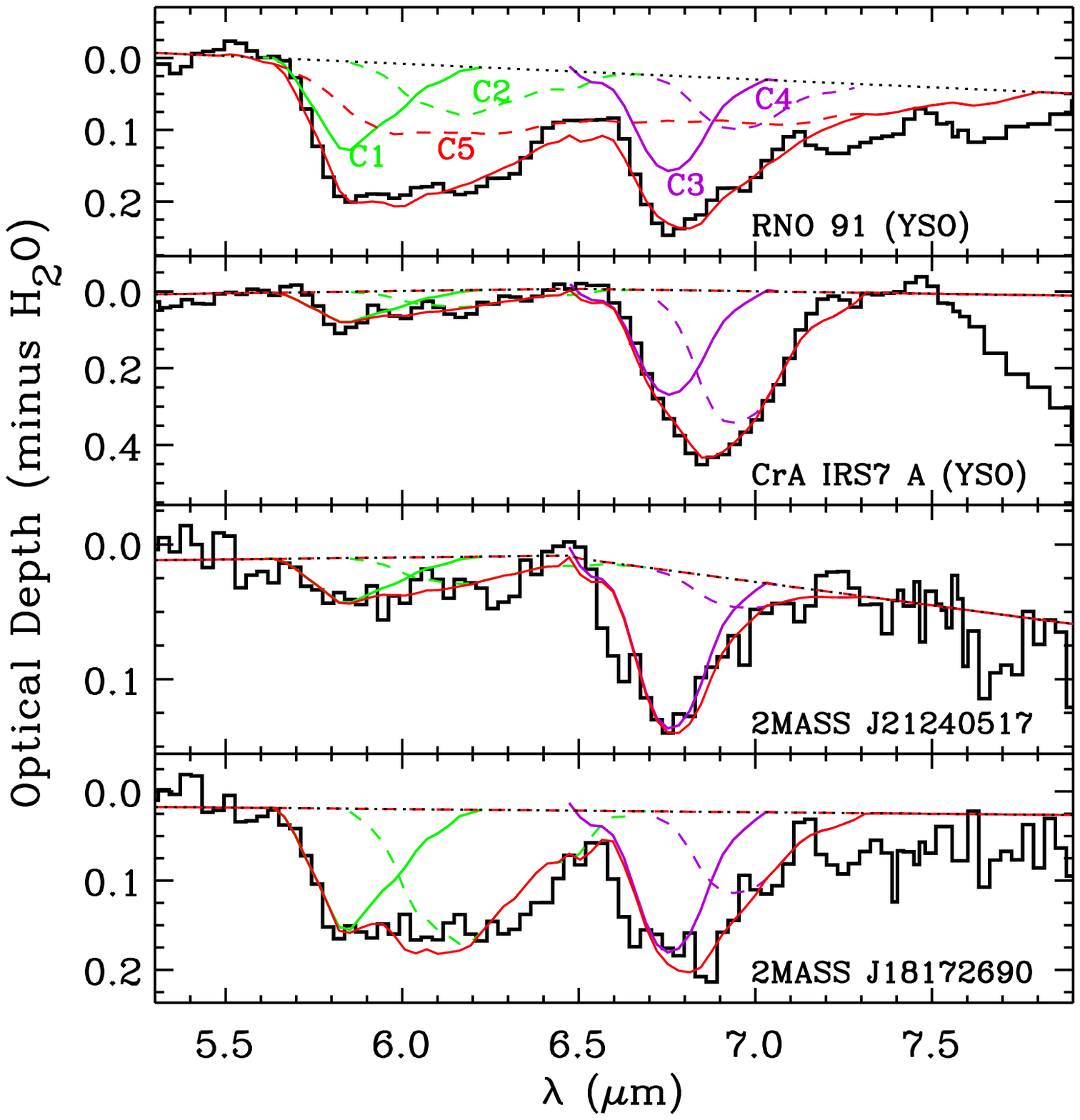}
\caption{Decomposition of the absorption features in the 5-8
  \mum\ spectra of two background stars (bottom panels), compared to
  two YSOs \citep{boo08} in the top panels. The components are labeled
  C1 (solid green), C2 (dashed green), C3 (solid purple), C4 (dashed
  purple), and C5 (dashed red). The sum of the components is indicated
  with a thick red line.  Note that the relative strengths of the
  components (e.g., C3/C1) varies strongly toward both YSOs and
  background stars. Component C5 has not been detected toward any
  background star. All components are added to a local straight line
  continuum (dotted black line) to remove small inaccuracies from the
  global continuum fitting process.}~\label{f:decomp}
\end{figure}

The continuum fits, as indicated with red lines in Fig.~\ref{f:obs1}
(left panels), but excluding modeled H$_2$O and silicate bands, were
used to put the spectra on an optical depth scale. The right panels of
Fig.~\ref{f:obs1} show that the optical depth in the 7.5-8.0
\mum\ region is not always fully explained by the modeled H$_2$O and
silicate bands. This may be related to inaccuracies in the applied
extinction curve and stellar model.  Therefore, in the subsequent
analysis of the optical depth spectra, local straight line continua
between 7.5 and 5.3 or 6.4 \mum\ were used to correct for these small
offsets (see Fig.~\ref{f:decomp} for examples).

Peak and integrated optical depths were determined for the 6.0 and
6.85~\mum\ bands, before and after the O-H bending mode of H$_2$O was
subtracted using an amorphous ice spectrum at $T$=10~K
(\citealt{hud93}; Table~\ref{t:tau}).  The features in the
H$_2$O-subtracted spectra were also decomposed into the five
components discussed in \citet{boo08}. These components (labeled
``C1''-``C5'') are useful in characterizing source-to-source
variations of the absorption in the 5-7 \mum\ wavelength region
(\S\ref{sec:h2o_c4c3}). They do not represent a unique decomposition,
however, and each component may be the result of absorption by more
than one molecular band.  Examples of this decomposition are shown in
Fig.~\ref{f:decomp}. For comparison two YSOs are shown as well
\citep{boo08}. Clearly, the relative depths of the components vary
considerably between background stars and between YSOs.

\begin{figure*}
\includegraphics[angle=90, scale=0.60]{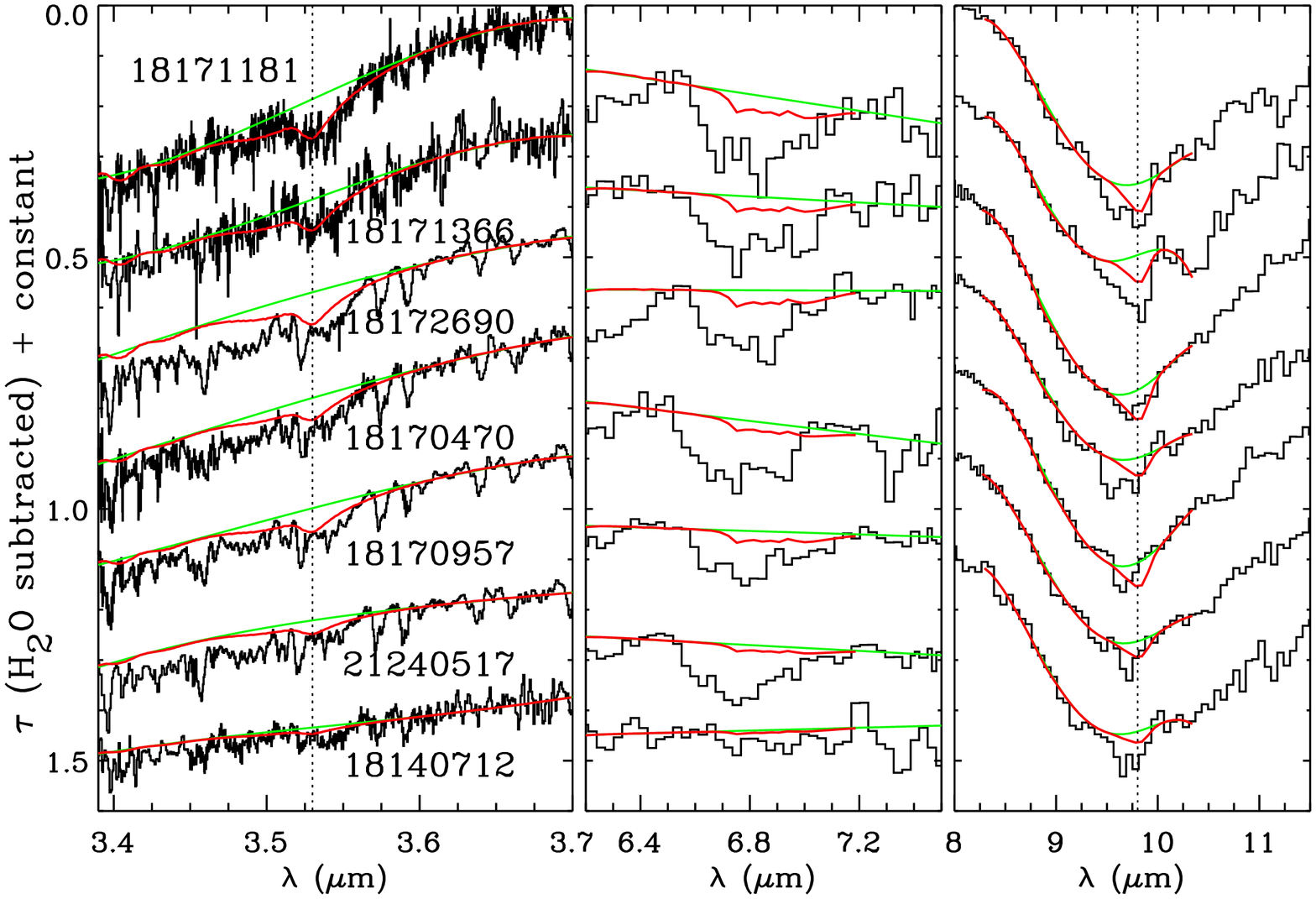}
\caption{{\bf Left panel:} Sources with detections (top 2 sources) or
  tentative detections of the C-H stretching mode of solid CH$_3$OH at
  3.53~\mum. The green line is the assumed local continuum and the red
  line the laboratory spectrum H$_2$O:CH$_3$OH:CO:NH$_3$=100:10:1:1
  ($T=10$ K; \citealt{hud93}) corresponding to the column densities
  listed in Table~\ref{t:colden}.  {\bf Middle panel:} the
  6.3-8.0~\mum\ region of the same sources as in the left panel
  showing the C-H bending mode of CH$_3$OH (red) with the assumed
  local continuum (green) at the same column densities as in the left
  panel.  The observed absorption is much deeper due to an overlapping
  band of (possibly) NH$_4^+$ (\S\ref{sec:col}). {\bf Right panel:}
  the 8-11.5~\mum\ region of the same sources as in the left panel
  showing the C-O stretching mode of CH$_3$OH (red) with the assumed
  local continuum (green) at the same column densities as in the left
  panel. Note that a scaling factor along the optical depth axis was
  applied for the spectra to fit in the panel.}~\label{f:ch3oh}
\end{figure*}

Column densities were determined for solid H$_2$O, \ammonium, and
CO$_2$ (Table~\ref{t:colden}), using the intrinsic integrated band
strengths listed in Table~\ref{t:avalues}.  The H$_2$O column density
is unreliable if no $L-$band spectrum is available (7 out of 31
sources), and the 13~\mum\ libration mode is the main tracer.  The
depth of the latter is strongly affected by the assumed shape of the
extinction curve and by overlapping stretching and bending modes of
silicates.  The \ammonium\ column density is derived, assuming that
the entire 6.85~\mum\ band (i.e., both components ``C3'' and ``C4''
listed in Table~\ref{t:tau}) can be attributed to it, after correction
for any overlapping bands of the C-H deformation mode of solid
CH$_3$OH (see below). This identification has not been fully
validated, however, because no other \ammonium\ bands have been
detected \citep{sch03, boo08}.  The identification of the
15.2~\mum\ band is not disputed, and the solid CO$_2$ column density
is listed for the 8 sources for which this wavelength range was
observed. All CO$_2$ observations were done at a low resolving power
of $\sim$90, which is insufficient to assess the composition or
thermal processing history of the ices \citep{pon08}.

The background star spectra were also searched for CH$_3$OH, HCOOH,
NH$_3$ and CH$_4$ species that are often detected toward YSOs.  For
solid CH$_3$OH, the 3.53~\mum\ C--H stretching and the 9.7~\mum\ C--O
stretching modes are usually used for column density determinations
\citep{boo08, bot10}. The 3.53~\mum\ band is preferred because it is
free from interfering strong bands. Nevertheless it overlaps with the
3.47~\mum\ band of hydrocarbons \citep{all92, bro96} or NH$_3$
hydrates \citep{dar01}, as well as several photospheric lines. Thus,
the most certain detections of solid CH$_3$OH are toward background
stars that show a {\it distinct} 3.53~\mum\ band. Fig.~\ref{f:ch3oh}
shows that this is the case for certainly two lines of sight:
2MASS~J18171181-0814012 and 2MASS~J18171366-0813188.  The strongest
detection (2MASS~J18171181-0814012) was verified by observations in
two different observing runs, using two different telluric standard
stars. A good agreement is found for the strength of the
9.7~\mum\ band. Five additional lines of sight likely show solid
CH$_3$OH as well, although in these cases the 3.53~\mum\ feature is
not as distinct from other fluctuations in the surrounding spectra
(Fig.~\ref{f:ch3oh}). For all of these 7 sources the CH$_3$OH
abundance relative to H$_2$O is in the 5-12\% range. A similar number
of sources have upper limits of comparable magnitude, and thus the
CH$_3$OH abundance varies in different environments.

For the YSOs studied in \citet{boo08}, HCOOH is considered a detection
only if the 7.25~\mum\ C-H deformation mode has been detected. This is
not the case for any of the background stars.  Based on the strength
of the C1 component (Table~\ref{t:tau}), which may mostly be due to
the C=O stretch mode of HCOOH, the abundance can in many lines of
sight be constrained to less than or comparable to the typical
detections toward YSOs of 2--5\% relative to H$_2$O, however
(Table~\ref{t:colden}).

Finally, for NH$_3$ and CH$_4$ the 8.9~\mum\ umbrella and
7.68~\mum\ bending modes are used to determine column density upper
limits.  For CH$_4$ the upper limits are sometimes comparable but
rarely less than the detections of $\sim$4\% toward YSOs
\citep{obe08}.  For NH$_3$ the upper limits are typically $\sim$20\%
relative to H$_2$O (Table~\ref{t:colden}), and are thus not
significant compared to the detections of 2-15\% toward YSOs
\citep{bot10}.

\begin{figure*}[b]
\includegraphics[angle=90, scale=0.70]{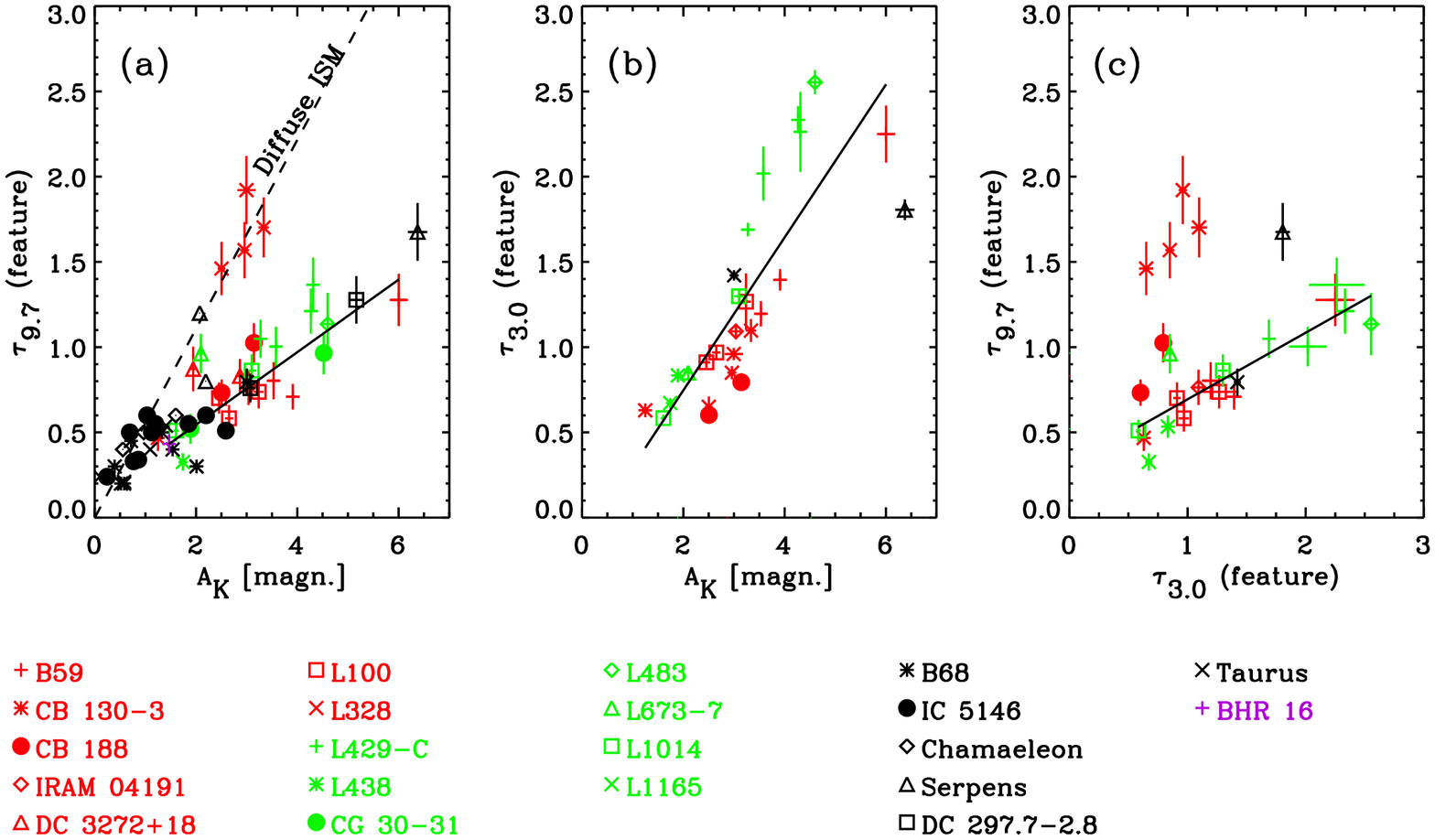}
\caption{Correlation plots of $A_{\rm K}$ with the peak optical depth
  of the 9.7~\mum\ feature {\bf (a)}, of $A_{\rm K}$ with the peak
  optical depth of the 3.0~\mum\ feature {\bf (b)}, and of the peak
  optical depths of the 3.0 and 9.7~\mum\ features {\bf
    (c)}. Background stars tracing different clouds and cores are
  indicated with different symbols and colors. Solid black lines show
  least-square fits to all data points, excluding the ones strongly
  affected by diffuse medium absorption (L328) and the ones in the
  Taurus and Serpens molecular clouds. The fit in panel {\bf (a)} was
  only done for $A_{\rm K}>1.4$ mag. The dashed line in that panel is
  the diffuse medium relation taken from the literature \citep{whi03},
  and the values for the B68, IC 5146, Chamaeleon, and most Serpens
  and Taurus sources were taken from \citet{chi07}.}~\label{f:corr1}
\end{figure*}

\begin{figure*}
\center
\includegraphics[angle=90, scale=0.50]{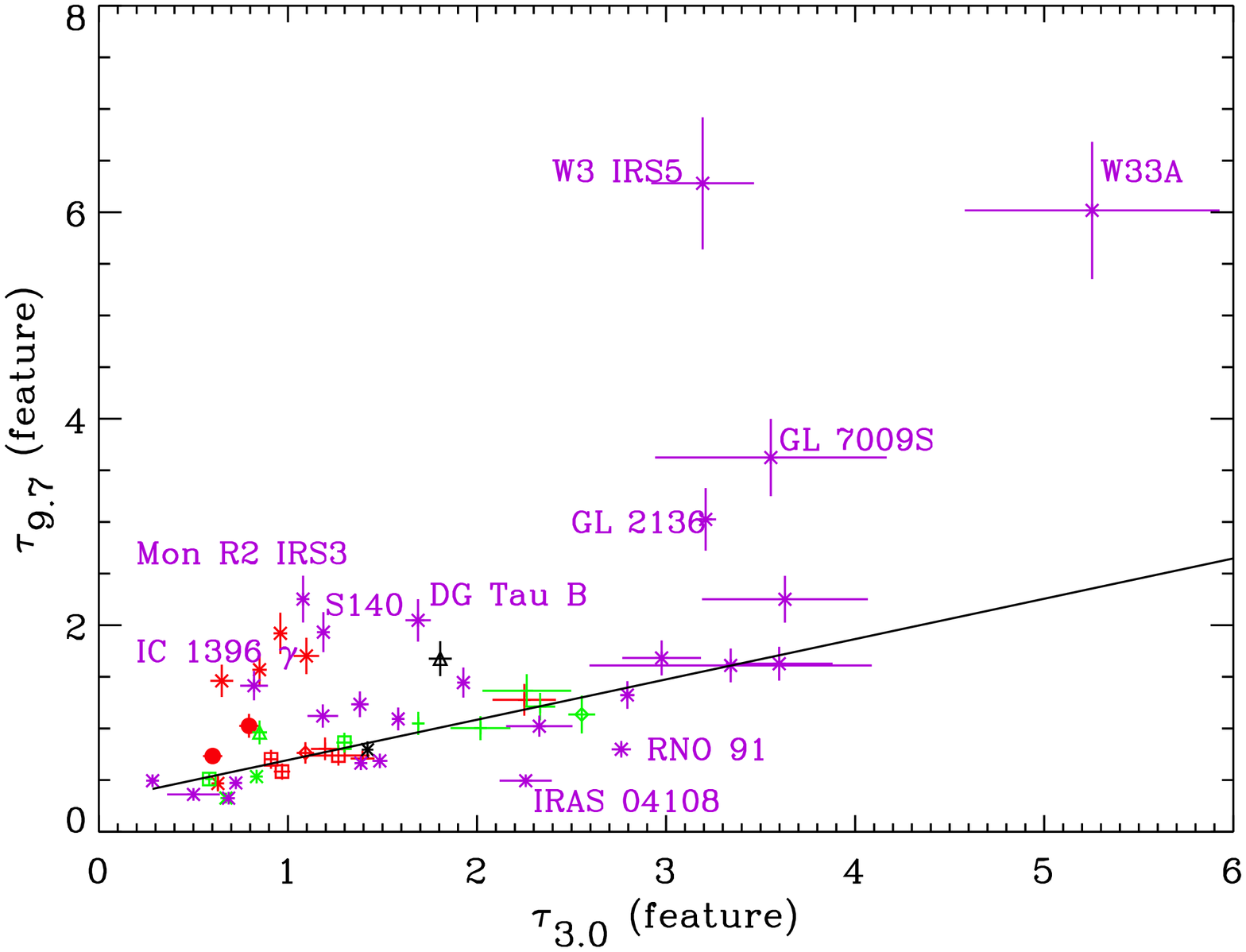}
\caption{Peak optical depths of the 3.0 and 9.7~\mum\ absorption
  features for both background stars (in black, red and green; see
  legend of Fig.~\ref{f:corr1}) and YSOs (purple asterisks). Only YSOs
  with reliable 3 \mum\ spectra are plotted. The solid line represents
  the fit for background stars as derived in Fig.~\ref{f:corr1} panel
  (c). YSOs deviating more than 3$\sigma$ from the solid line are
  labeled, except for IC 1396 $\alpha$ and B5 IRS1 (both above the
  solid line), and IRAS 03439, L1448 IRS1 (below the solid line),
  which are not labeled to avoid crowding in the
  plot.}~\label{f:corr1.5}
\end{figure*}

\subsection{Correlation Plots}~\label{sec:60}

To facilitate the analysis of the observational parameters in many
sight-lines in many different environments (quiescent core and cloud
media, low mass YSOs, massive YSOs), a number of correlation plots are
presented below.  The values for the YSOs were taken from
\citet{boo08} and \citet{rea09}.  Note that in plots involving $A_{\rm
  K}$ no YSOs are present because of the unknown shape of the
intrinsic K-band continuum emission. Also, background stars with
$\tau_{3.0}$ upper limits have been excluded from the plots.

\subsubsection{$\tau_{9.7}$ versus $A_{\rm K}$ }~\label{sec:tau97_ak}

The plot of the peak optical depth of the 9.7~\mum\ band of silicates
against K-band continuum extinction $A_{\rm K}$ shows an interesting
behavior (Fig.~\ref{f:corr1}a). While at low $A_{\rm K}$ the diffuse
medium relation \citep{whi03}

\begin{equation}
\tau_{9.7}=0.554\times A_{\rm K}~\label{eq:tau97akdism}
\end{equation}

\noindent is followed (assuming $A_{\rm V}/A_{\rm K}$=8.8), a less
steep relation is observed at $A_{\rm K}\geq$1 mag in dense
sight-lines. This was discovered by \citet{chi07} in the IC 5146, B68
and B59 cores and Serpens and Taurus clouds, and here this behavior is
extended in more lines of sight, at higher K-band extinctions. Some
background stars, however, follow the diffuse medium correlation, most
notably those behind L328, and, as noted by \citet{chi07}, one source
behind Serpens (2MASS~J18285266+0028242).  Thus, in these lines of
sight the extinction may well be dominated by diffuse rather than
dense dust. Excluding these particular background stars, a polynomial
was fitted to objects with $A_{\rm K}>$1.4 mag:

\begin{equation}
\tau_{9.7}=(0.12\pm 0.05)+(0.21\pm 0.02)\times A_{\rm
  K}~\label{eq:tau97ak}
\end{equation}

\noindent It is worth noting that for some cores all background stars
lie systematically above or below the fit, e.g., L429-c sources lie
above it and B59 sources below it.

\subsubsection{$\tau_{3.0}$ versus $A_{\rm K}$ }~\label{sec:tau30_ak}

The peak optical depth of the 3.0~\mum\ H$_2$O ice band correlates
strongly with $A_{\rm K}$ (Fig.~\ref{f:corr1}b). There is no
indication of flattening. A linear fit to all data points yields:

\begin{equation}
\tau_{3.0}=(-0.15\pm 0.13)+(0.45\pm 0.05)\times A_{\rm
  K}~\label{eq:tau30ak}
\end{equation}

\noindent This relation implies a $\tau_{3.0}=0$ cut-off value of
$A_{\rm K}=0.33\pm 0.30$, which corresponds to $A_{\rm V}=2.9\pm 2.6$
(assuming $A_{\rm V}/A_{\rm K}$=8.8). This is the so-called ice
formation threshold, and it is comparable to the value observed for
the Taurus Molecular Cloud ($A_{\rm V}=3.1\pm 0.6$; \citealt{chi95}),
but smaller than the $A_{\rm V}=10-15$ mag quoted by \citet{tan90} for
Ophiuchus. The large uncertainty must reflect different ice formation
thresholds in the different environments traced by the observations,
or the presence of different amounts of ice-less diffuse dust along
the line of sight.  For example, all L328, B59 and CB 188 sources lie
to the right of the fitted line. For L328 this most likely indicates a
significant contribution of extinction in diffuse foreground clouds as
indicated by the $\tau_{9.7}$ versus ${\rm A_K}$ relation
(\S\ref{sec:tau97_ak}).  On the other hand, all L429-C sources lie to
the left of the fit, indicating a low ice formation threshold.

\subsubsection{$\tau_{3.0}$ versus $\tau_{9.7}$}~\label{sec:tau30_97}

Although the general relation of $\tau_{3.0}$ with $\tau_{9.7}$ is
evident from panels (a) and (b) of Fig.~\ref{f:corr1}, it is plotted
separately in panel (c). A linear fit to all points except those from
L328 (\S\ref{sec:tau97_ak}) yields

\begin{equation}
\tau_{9.7}=(0.36\pm 0.09)+(0.36\pm 0.06)\times
\tau_{3.0}~\label{eq:tau3097}
\end{equation}

\noindent For this particular relation, values for the YSOs are
available from \citet{boo08}. They are displayed in
Fig.~\ref{f:corr1.5}, where the linear fit from Eq.~\ref{eq:tau3097}
above has been plotted.  YSOs deviating more than 3$\sigma$ from the
fit are labeled.  YSOs lying below the line may have some of the
silicate band filled in by emission, while sources above the line may
have significant amounts of warm dust in which the H$_2$O ices have
evaporated. Interestingly, the latter are mostly the massive YSOs and
YSOs that show a ``red'' 6.85~\mum\ band, indicative of thermally
processed ices (\S\ref{sec:h2o_c4c3}).

\begin{figure}
\includegraphics[angle=90, scale=0.36]{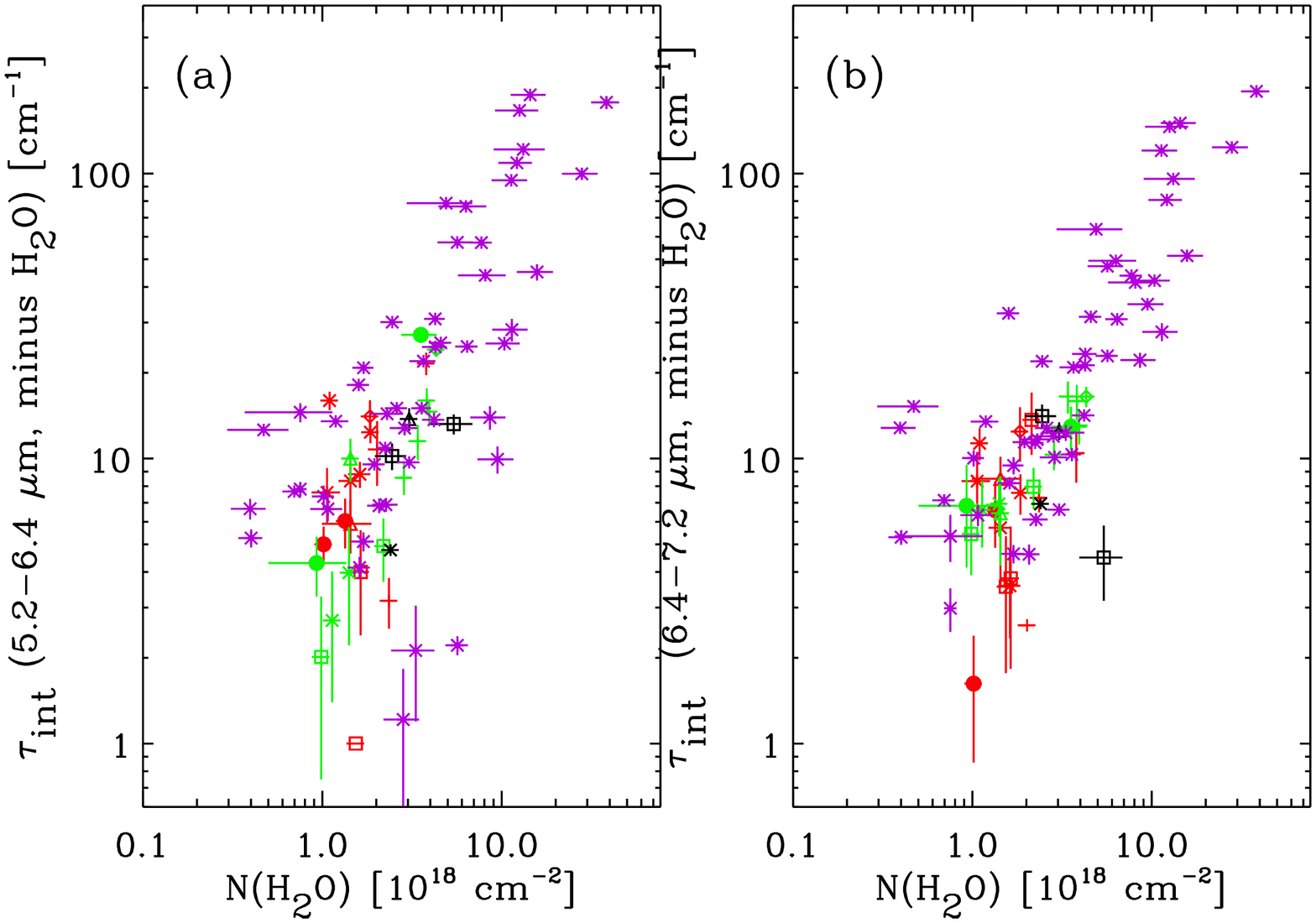}
\caption{Integrated optical depths of H$_2$O-subtracted spectra in the
  5.2-6.4 ({\bf panel a}) and 6.4-7.2~\mum\ ({\bf panel b}) ranges
  plotted against the H$_2$O ice column densities.  The symbol types
  and colors have the same meaning as in Figs.~\ref{f:corr1}
  and~\ref{f:corr1.5}. Despite the large scatter, it can be concluded
  that the background stars follow the same correlations as the YSOs
  (purple asterisks), i.e., the carriers of the absorption are already
  present in quiescent cloud material.}~\label{f:corr2}
\end{figure}

\subsubsection{6.0 $\mu$m band versus $N({\rm H_2O})$}~\label{sec:60_h2o}

Previous work by \citet{kea01} toward massive YSOs, \citet{kne05}
toward background stars and \citet{boo08} toward low mass YSOs has
shown that the 6.0~\mum\ band can only partly be attributed to solid
H$_2$O. Although H$_2$O mixed with realistic concentrations of CO$_2$
fits the long wavelength side (component ``C2'' in \citealt{boo08})
better than pure H$_2$O \citep{kne05,obe07}, it does not explain all
absorption.  This is the case for the background stars studied here as
well.  To quantify this, the integrated absorption remaining after
subtraction of an amorphous, pure H$_2$O laboratory ice at $T=$10~K
\citep{hud93} is plotted against $N{\rm (H_2O)}$, showing that the
background stars are in line with the YSOs (Fig.~\ref{f:corr2}a;
Table~\ref{t:tau}).  The scatter is large, however, which is reflected
in a linear correlation coefficient of 0.79. Some of this might be due
to uncertainties in $N{\rm (H_2O)}$, as many YSOs do not have reliable
3.0~\mum\ spectra available. Indeed, when only including sources with
reliable 3.0~\mum\ spectra, the correlation coefficient increases to
0.85.  Another source of scatter may be the inaccuracy in the
continuum determination, especially for the YSOs. Intrinsic abundance
variations of the different carriers of the 5.2-6.4~\mum\ ``excess''
absorption relative to H$_2$O must cause some of the scatter as well
(Fig.~\ref{f:decomp}).

\subsubsection{6.85 $\mu$m band versus $N({\rm H_2O})$}~\label{sec:68_h2o}

The 6.85~\mum\ band is detected toward many background stars.  Its
integrated optical depth, calculated after subtracting the long
wavelength wing of the H$_2$O bending mode, is correlated with $N{\rm
  (H_2O)}$ in Fig.~\ref{f:corr2}b, where the YSOs have been plotted as
well. As for the 6.0~\mum\ band, the scatter is large.  At a given
value of $N{\rm (H_2O)}$, the 6.85~\mum\ band depth varies by a factor
of 2. The linear correlation coefficient is 0.88 when including all
background stars and YSOs.  Within the scatter, there is no systematic
difference in the strength of the 6.85~\mum\ band between YSOs and
background stars.

\subsubsection{C4/C3 Ratio versus H$_2$O abundance }~\label{sec:h2o_c4c3}

\begin{figure}
\includegraphics[angle=90, scale=0.41]{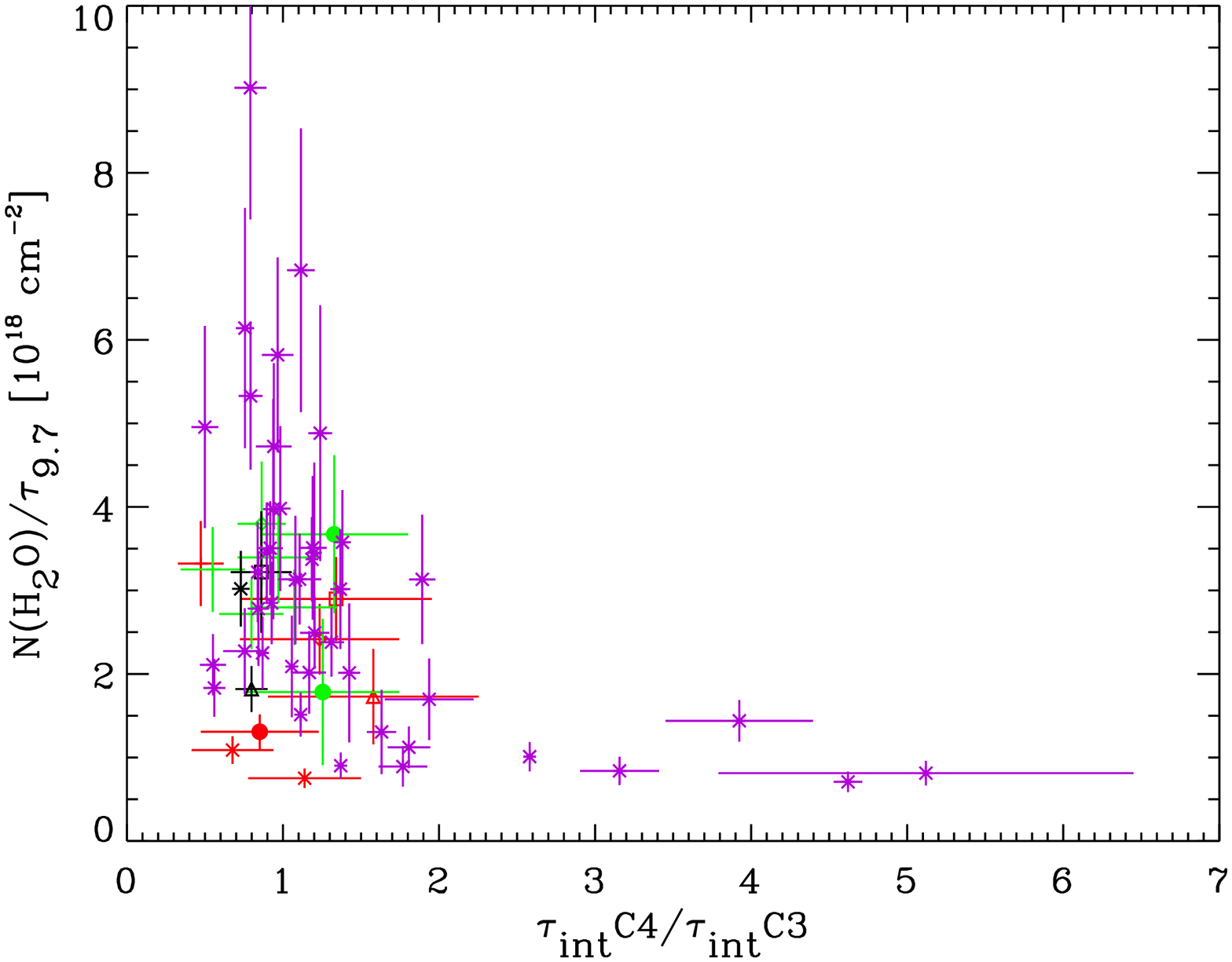}
\caption{Ratio of integrated optical depths of the C4 and C3
  components (long and short wavelength components of the
  6.85~\mum\ band) plotted versus the H$_2$O ice column density
  normalized to the peak optical depth of the 9.7~\mum\ band (a
  measure of the H$_2$O ice abundance). Only values with significance
  $>2\sigma$ are plotted. Different colors and symbols refer to
  background stars behind different clouds and cores (see legend of
  Fig.~\ref{f:corr1}). The values for the YSOs, indicated with purple
  asterisks, are taken from \citet{boo08}.  While nearly all sources
  have C4/C3 ratios near 1, a number of YSOs have significantly larger
  ratios and at the same time low H$_2$O abundances.
}~\label{f:corr3}
\end{figure}

The 6.0 and 6.85~\mum\ bands were decomposed into components C1-C5 in
\citet{boo08}. It was found that the C4/C3 component ratio varies
significantly between YSOs. Sources with large C4/C3 ratios tend to
have low H$_2$O ice abundances (represented by the $N{\rm
  (H_2O)}/\tau{_{9.7}}$ ratio) and large degrees of ice
crystallization and segregation \citep{kea01, boo08}, thus indicating
that the C4/C3 ratio increases as a result of thermal processing.  A
similar decomposition was done for the background stars
(Fig.~\ref{f:corr3}), and indeed they have low C4/C3 ratios as
expected for the cold, unprocessed sight-lines that they trace.
Nevertheless, some of the background stars (e.g., toward L328) have
low H$_2$O ice abundances, likely because of the presence of diffuse,
ice-free, foreground clouds (\S\ref{sec:tau97_ak}) rather than because
of the effects of ice sublimation as for YSOs.

\section{Discussion}~\label{sec:dis}

\subsection{Ice Abundances}~\label{sec:dis1}

The infrared spectra of ices and dust toward background stars
presented here provide a baseline for the astrochemical evolution of
YSOs.  It is shown that the strength and shape of the 3.0, 6.0, 6.85
and 15~\mum\ ice bands are similar for the quiescent material in
isolated cores traced by background stars compared to most previously
studied YSOs.  Correspondingly, the relative abundances of the ices
with secure (CO$_2$, CH$_3$OH) and less secure identifications
(NH$^+_4$) are similar to most YSOs.  Species responsible for the 6.0
\mum\ band, besides H$_2$O (probably HCOOH, H$_2$CO, and NH$_3$), must
have abundances similar to YSOs as well.  This confirms earlier work
on smaller samples \citep{kne05, ber05}, and it firmly establishes
that most ices are formed early on in the molecular cloud evolution,
and are largely unaltered during the further evolution of the region
(e.g., the star formation process) until the YSO thermally processes
the ices in the envelope.

The two background stars with the most secure CH$_3$OH detections
(2MASS~J18171181-0814012 and 2MASS~J18171366-0813188), at $\sim$11\%
with respect to H$_2$O, are both located behind the core L429-C.
These lines of sight show some of the deepest ice and dust features in
the sample, which may have facilitated the detection of the weak 3.53
\mum\ CH$_3$OH feature.  Indeed, many other background stars in the
sample have upper limits or tentative detections at abundances not
much less (6-10\%) than the detections.  Some,
e.g. 2MASS~J04393886+2611266 in the Taurus Molecular Cloud, have
tighter upper limits of 1-3\% (\citealt{chi96}; Table~\ref{t:colden}),
however.  Large CH$_3$OH abundance variations are also observed toward
YSOs \citep{dar99, pon03b, boo08}.

The presence of solid CH$_3$OH in quiescent clouds is not unexpected,
following theoretical studies \citep{tie87, kea01a}, laboratory
studies \citep{wat02, fuc09} and time-dependent (Monte Carlo)
simulations \citep{cup09}. While the laboratory experiments have not
yet been fully translated into kinetic parameters, in a general sense,
the theoretical studies and kinetic Monte Carlo simulations do show
that the CH$_3$OH/CO ratio is very sensitive to the dust temperature
and to the ratio of atomic H to CO in the accreting gas \citep{kea01a,
  cup09}. In this model, an H$_2$O-rich mantle forms first, because of
its higher sublimation temperature, and at $T<$17~K a CO mantle forms
on top of that. Accreting H atoms then form H$_2$CO and CH$_3$OH.
High CH$_3$OH/CO ratios are indicative of high H/CO ratios of the
accreting gas (due to relatively low densities or high local cosmic
ray fluxes that destroy H$_2$).  Alternatively, low CH$_3$OH/CO ratios
may reflect relatively elevated dust temperatures ($T>15$ K;
\citealt{cup09}).  Thus local physical conditions are a dominating
factor in the formation of CH$_3$OH, and this may explain the large
observed CH$_3$OH abundance variations in different lines of sight.

\subsection{Processing of the Ices}~\label{sec:dis2}

\begin{figure}
\includegraphics[angle=90, scale=0.50]{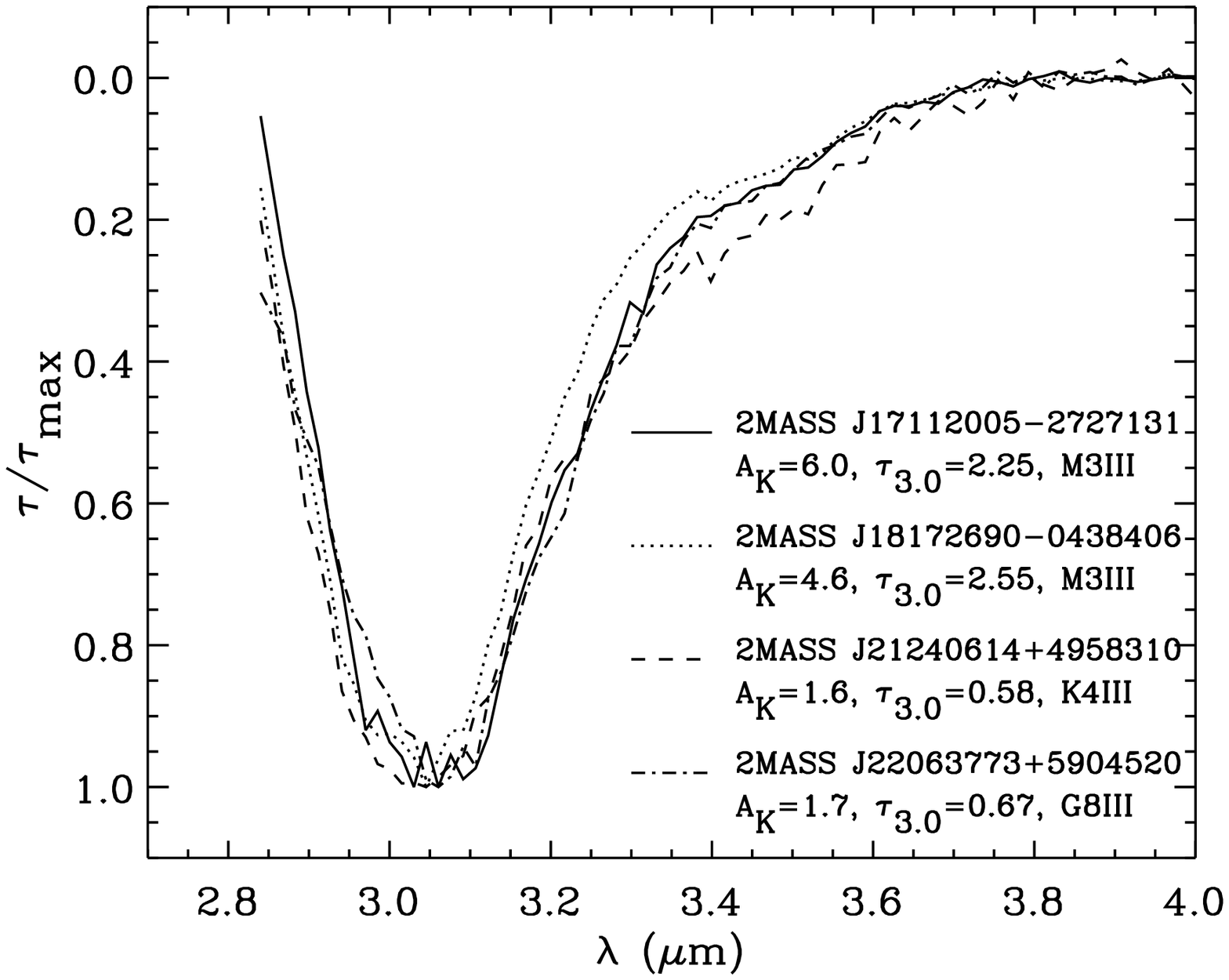}
\caption{Profiles of the 3 \mum\ ice bands observed toward two
  background stars with the highest (solid and dotted lines), and two
  background stars with the lowest (dashed and dash-dotted lines)
  extinctions and peak optical depths in the sample. The spectra are
  smoothed to a resolving power of 100. The profiles are very similar,
  indicating a lack of grain growth.}~\label{f:3um}
\end{figure}

Ice evolution is observed in the envelopes of some YSOs, and it
appears that this is mostly due to the effects of heating: apolar
CO-rich ices sublimate, and H$_2$O-rich ices crystallize \citep{smi89,
  ger99, boo00, pon08, boo08}.  The carrier of the 6.85~\mum\ band is
affected by heating as well: it shifts to longer wavelengths
\citep{kea01, boo08}.  Toward background stars, none of these effects
are observed.  Fig.~\ref{f:corr3} shows that all background stars with
high quality spectra have 6.85~\mum\ bands peaking at short
wavelengths (low C4/C3 ratios). The observed 3.0 \mum\ bands
(Fig.~\ref{f:3um}) are characteristic of low temperature, amorphous
ices. Crystalline ices would peak near 3.1 \mum\ \citep{smi89}.

Finally, an absorption feature in the 5-8~\mum\ region referred to as
C5 in \citet{boo08} (see also \citealt{gib02}), has not been observed
toward background stars (Table~\ref{t:tau}). Although continuum
reliability is generally a problem for this shallow feature,
detections ($>3\sigma$) were claimed in $\sim$ 30\% of the low mass
YSOs and $\sim$ 50\% of the massive YSOs.  The feature could be due to
the formation of new molecules by energetic processing of the ices,
possibly hydrocarbons \citep{gre95} or ions \citep{sch03}. Its absence
towards background stars is consistent with an origin in ices
processed in the YSO environment, and not by external cosmic rays.

\subsection{Ices, Silicates, and the Extinction Curve}~\label{sec:dis3}

Observations show that the properties of dust in diffuse and dense
clouds are different. In contrast to diffuse clouds, the mid-infrared
extinction in the dense cores studied here ($A_{\rm K}>$1 magn) stays
high: at 25 \mum\ it is still half of that at 2
\mum\ (Fig.~\ref{f:ext}b; \citealt{cha09, mcc09}). This is most likely
caused by grain growth.  Indeed, the calculated $R_{\rm V}$=5.5
(``case B'') extinction curve from \citet{wei01} is closer to the
observations than their $R_{\rm V}$=3.1 ``case A'' model
(Fig.~\ref{f:ext}c; \citealt{cha09, mcc09}).  Nevertheless, even this
extreme case of $R_{\rm V}$=5.5 is systematically too shallow at
wavelengths above 5 \mum, while the silicate features are too deep.
The present work shows that the contribution of ices, which are not
included in the \citet{wei01} models, to the extinction is not very
large: $\sim$25\% in the 11-17 \mum\ wavelength region.  Instead, a
model with larger grain sizes is needed.  Such model would also need
to take into account the observed weakness of the 9.7 \mum\ band of
silicates (Figs.~\ref{f:ext}c and~\ref{f:corr1}a).

Despite the need for larger grains in dense clouds, compared to
diffuse clouds, there is no evidence for progressive grain growth in
dense clouds for $A_{\rm K}> 1$ magn.  The empirical extinction curve
(Fig.~\ref{f:ext}) satisfactorily fits most background stars in the
sample, for the range of observed $A_{\rm K}$ values of 1.3-6 mag.
The same conclusion can be drawn from Fig. 1 in \citet{mcc09}.  Also,
the 3.0 \mum\ band profiles are very similar over this $A_{\rm K}$
range (Fig.~\ref{f:3um}), while a broadening and shift to longer
wavelengths would be expected if grain growth occurred.  Similarly,
the observed linear relation of the 3.0 \mum\ band with dust column
density indicators such as the 9.7~\mum\ band of silicates and $A_{\rm
  K}$ (Fig.~\ref{f:corr1}) indicates a lack of progressive grain
growth, since larger grains produce 3.0 \mum\ bands with smaller peak
optical depths.

Recent work shows that the silicate band profiles are inconsistent
with grain growth beyond 1 \mum\ \citep{bre11}, similar to what is
found for the 3.0 \mum\ ice band \citep{smi89}.  It is thus unlikely
that grain growth can fully explain the shallower $A_{\rm K}$ versus
$\tau_{9.7}$ relation in the dense medium compared to the diffuse
medium (Fig.~\ref{f:corr1}a). Different dust compositions may play a
role as well. The silicate bands in dense clouds show a short
wavelength wing that could be due to a larger pyroxene/olivine
abundance ratio (\S\ref{sec:cont}).  For a more thorough investigation
of the $A_{\rm K}$ versus $\tau_{9.7}$ relation one is referred to
\citet{bre11}.

\section{Conclusions and Future Work}~\label{sec:concl}

The analysis of 2-25~\mum\ photometry and spectra of 31 stars tracing
the quiescent medium in 16 isolated dense cores yields the following
conclusions:

\begin{enumerate}

\item An empirical ``high resolution'' extinction curve derived from one
  of the background stars confirms extinction curves derived from broad
  band photometry in high density clouds \citep{cha09}. The curve
  remains remarkably flat above 10~\mum, such that at 25~\mum\ the
  extinction is still half that at 2.2~\mum. The main caveat is that
  it is assumed the near-infrared curve is the same as that derived by
  \citet{ind05}.

\item The empirical extinction curve, in combination with model spectra
  of giants, and absorption spectra of H$_2$O ice and silicates is
  able to fit the observed data of most background stars well. No
  systematic deviations are seen as a function of core or $A_{\rm K}$
  values (noting that the sample includes only high extinction lines
  of sight with $A_{\rm K}\gtrsim 1.5$ mag).

\item Model fits of the attenuated stellar continuum are used to
  isolate the ice and dust absorption features. The 9.7~\mum\ band of
  silicates shows a shallower relation with $A_{\rm K}$ compared to
  the diffuse interstellar medium, thus confirming the results of
  \citet{chi07} in many more lines of sight. Some background stars,
  e.g., those behind the L328 core, follow the diffuse medium relation,
  however.

\item The 3.0~\mum\ band of H$_2$O ice is detected in all observed
  sources. A strong correlation is observed between its peak optical
  depth and $A_{\rm K}$.  Its shape is independent of $A_{\rm K}$,
  indicating a lack of grain growth at high $A_{\rm K}$ values.  This
  may imply that the shallower relation of $\tau_{9.7}$ with $A_{\rm
    K}$ in the dense ISM compared to the diffuse ISM is not caused by
  grain growth. Possibly the dust composition is different, as also
  suggested by the broader 9.7~\mum\ band in dense sight-lines (both
  YSOs and background stars).

\item The 6.0 and 6.85~\mum\ absorption bands are detected in most
  background stars. As for YSOs, the 6.0~\mum\ band is only partially
  explained by the bending mode of H$_2$O ice.  Additional species,
  such as HCOOH, H$_2$CO, CH$_3$OH, and NH$_3$ must be responsible for
  this.  Together with the consistent presence of the 6.85~\mum\ band,
  which may be due to NH$_4^+$, this shows that there is little
  difference between the ices in quiescent core regions and those
  surrounding most YSOs. Most ices are apparently formed in the
  quiescent cloud phase.

\item The discovery of solid CH$_3$OH in the L-band spectra of several
  background stars strengthens the previous point. It confirms recent
  models by \citet{cup09} that CH$_3$OH is efficiently formed on the
  grains in low temperature ($T<15$ K) clouds. Tight upper limits in
  previously studied sight-lines indicate that the ice composition
  varies in different environments at the onset of star formation,
  however. For a proper comparison with the models, observed
  CH$_3$OH/CO ratios are needed, i.e., more observations of solid CO
  toward background stars.

\item Signatures that are assigned to processing of the ices
  surrounding YSOs, such as profile changes in the 3 \mum\ ice band,
  the shifted 6.85~\mum\ band (i.e., a large C4/C3 ratio) and the
  presence of the broad C5 components, are not found toward the
  background stars.

\end{enumerate}

The present work studies ices and the extinction curve in the
quiescent medium in isolated dense cores.  YSOs in these cores are not
well studied.  On the other hand, the ices are well studied toward
YSOs in large clouds (Taurus, Serpens, Oph, Perseus; \citealt{boo08,
  pon08, obe08, bot10}), but the background star samples are small
\citep{kne05}.  Thus, future studies will need to compare YSOs and
background stars in the same environments to address the origin of the
observed variations of the CH$_3$OH abundance and the strength of the
6.0 and 6.8 \mum\ ice bands (Figs.~\ref{f:decomp} and~\ref{f:corr2}).
Furthermore, while there is now much observational evidence that the
dust properties in the dense and diffuse medium are quite different,
the cause of this is not understood, warranting further theoretical
and perhaps laboratory investigations.

\acknowledgments 

We thank the anonymous referee for comments that helped with improving
the presentation and conclusions of this work.  T.L.H. acknowledges
support for this work provided by NASA through contract 1316720 issued
by JPL/Caltech.  A.M.C. acknowledges support from the NASA
Astrobiology Institute (grant NNA09DA80A) and the IPAC Visiting
Graduate Student Fellowship Program.  This publication makes use of
data products from the Two Micron All Sky Survey, which is a joint
project of the University of Massachusetts and the Infrared Processing
and Analysis Center/California Institute of Technology, funded by the
National Aeronautics and Space Administration and the National Science
Foundation.


\begin{thebibliography}{}


\bibitem[Allamandola et al.(1992)]{all92} Allamandola, L.~J.,
  Sandford, S.~A., Tielens, A.~G.~G.~M., \& Herbst, T.~M.\ 1992, \apj,
  399, 134

\bibitem[Beichman et al.(1986)]{bei86} Beichman, C.~A., Myers, P.~C.,
  Emerson, J.~P., Harris, S., Mathieu, R., Benson, P.~J., \& Jennings,
  R.~E.\ 1986, \apj, 307, 337

\bibitem[Bergin et al.(2005)]{ber05} Bergin, E.~A., Melnick, G.~J.,
  Gerakines, P.~A., Neufeld, D.~A., \& Whittet, D.~C.~B.\ 2005, \apjl,
  627, L33

\bibitem[Bernstein et al.(2002)]{ber02} Bernstein, M.~P., Dworkin,
  J.~P., Sandford, S.~A., Cooper, G.~W., \& Allamandola, L.~J.\ 2002,
  \nat, 416, 401

\bibitem[Bohlin et al.(1978)]{boh78} Bohlin, R.~C., Savage, B.~D., \&
  Drake, J.~F.\ 1978, \apj, 224, 132

\bibitem[Bohren \& Huffman (1983)]{boh83} Bohren, C. F., \& Huffman,
  D.  R. 1983, Absorption and Scattering of Light by Small Particles
  (New York: John Wiley \& Sons)

\bibitem[Boogert et al.(1997)]{boo97} Boogert, A.~C.~A., Schutte,
  W.~A., Helmich, F.~P., Tielens, A.~G.~G.~M., \& Wooden, D.~H.\ 1997,
  \aap, 317, 929

\bibitem[Boogert et al.(2000)]{boo00} Boogert, A.~C.~A., et al.\ 2000,
  \aap, 353, 349

\bibitem[Boogert et al.(2002)]{boo02} Boogert, A.~C.~A., Blake, G.~A.,
  \& Tielens, A.~G.~G.~M.\ 2002, \apj, 577, 271

\bibitem[Boogert \& Ehrenfreund(2004)]{boo04} Boogert, A.~C.~A., \&
  Ehrenfreund, P.\ 2004, Astrophysics of Dust, 309, 547

\bibitem[Boogert et al.(2008)]{boo08} Boogert, A.~C.~A., et al.\ 2008,
  \apj, 678, 985

\bibitem[Bottinelli et al.(2010)]{bot10} Bottinelli, S., et al.\ 2010,
  \apj, 718, 1100

\bibitem[Brooke et al.(1996)]{bro96} Brooke, T.~Y., Sellgren, K., \&
  Smith, R.~G.\ 1996, \apj, 459, 209

\bibitem[Chapman et al.(2009)]{cha09} Chapman, N.~L., Mundy, L.~G.,
  Lai, S.-P., \& Evans, N.~J.\ 2009, \apj, 690, 496

\bibitem[Chiar et al.(1995)]{chi95} Chiar, J.~E., Adamson, A.~J.,
  Kerr, T.~H., \& Whittet, D.~C.~B.\ 1995, \apj, 455, 234

\bibitem[Chiar et al.(1996)]{chi96} Chiar, J.~E., Adamson, A.~J., \&
  Whittet, D.~C.~B.\ 1996, \apj, 472, 665

\bibitem[Chiar et al.(2007)]{chi07} Chiar, J.~E., et al.\ 2007, \apjl,
  666, L73

\bibitem[Chiar et al.(2011)]{chi11} Chiar, J.~E., et al.\ 2011, \apj
  (submitted)

\bibitem[Cuppen et al.(2009)]{cup09} Cuppen, H.~M., van Dishoeck,
  E.~F., Herbst, E., \& Tielens, A.~G.~G.~M.\ 2009, \aap, 508, 275

\bibitem[Dartois et al.(1999)]{dar99} Dartois, E., Schutte, W.,
  Geballe, T.~R., Demyk, K., Ehrenfreund, P., \& d'Hendecourt,
  L.\ 1999, \aap, 342, L32

\bibitem[Dartois \& d'Hendecourt(2001)]{dar01} Dartois, E., \&
  d'Hendecourt, L.\ 2001, \aap, 365, 144

\bibitem[Decin et al.(2004)]{dec04} Decin, L., Morris, P.~W.,
  Appleton, P.~N., Charmandaris, V., Armus, L., \& Houck, J.~R.\ 2004,
  \apjs, 154, 408

\bibitem[Dorschner et al.(1995)]{dor95} Dorschner, J., Begemann, B.,
  Henning, T., Jaeger, C., \& Mutschke, H.\ 1995, \aap, 300, 503

\bibitem[Evans(1999)]{eva99} Evans, N.~J., II 1999, \araa, 37, 311

\bibitem[Evans et al.(2003)]{eva03} Evans, N.~J., II, et al.\ 2003,
  \pasp, 115, 965

\bibitem[Evans et al.(2007)]{eva07} Evans, N.~J., II, et al.\ 2007,
  Final Delivery of Data from the c2d Legacy Project: IRAC and MIPS
  (Pasadena: SSC), http://ssc.spitzer.caltech.edu/legacy/

\bibitem[Fuchs et al.(2009)]{fuc09} Fuchs, G.~W., Cuppen, H.~M.,
  Ioppolo, S., Romanzin, C., Bisschop, S.~E., Andersson, S., van
  Dishoeck, E.~F., \& Linnartz, H.\ 2009, \aap, 505, 629

\bibitem[Gerakines et al.(1995)]{ger95} Gerakines, P.~A., Schutte,
  W.~A., Greenberg, J.~M., \& van Dishoeck, E.~F.\ 1995, \aap, 296,
  810

\bibitem[Gerakines et al.(1999)]{ger99} Gerakines, P.~A., et
  al.\ 1999, \apj, 522, 357

\bibitem[Gibb \& Whittet(2002)]{gib02} Gibb, E.~L., \& Whittet,
  D.~C.~B.\ 2002, \apjl, 566, L113

\bibitem[Gibb et al.(2004)]{gib04} Gibb, E.~L., Whittet, D.~C.~B.,
  Boogert, A.~C.~A., \& Tielens, A.~G.~G.~M.\ 2004, \apjs, 151, 35

\bibitem[Greenberg et al.(1995)]{gre95} Greenberg, J.~M., Li, A.,
  Mendoza-Gomez, C.~X., Schutte, W.~A., Gerakines, P.~A., \& de Groot,
  M.\ 1995, \apjl, 455, L177

\bibitem[Hagen et al.(1981)]{hag81} Hagen, W., Tielens, A.~G.~G.~M.,
  \& Greenberg, J.~M.\ 1981, Chem. Phys., 56, 367

\bibitem[Hudgins et al.(1993)]{hud93} Hudgins, D.~M., Sandford, S.~A.,
  Allamandola, L.~J., \& Tielens, A.~G.~G.~M.\ 1993, \apjs, 86, 713

\bibitem[Indebetouw et al.(2005)]{ind05} Indebetouw, R., et al.\ 2005,
  \apj, 619, 931

\bibitem[Kane \& Sahu(2003)]{kan03} Kane, S.~R., \& Sahu, K.~C.\ 2003,
  \apj, 582, 743

\bibitem[Keane(2001a)]{kea01a} Keane, J.~V.\ 2001a, Ph.D.~Thesis,
  Chapter 7

\bibitem[Keane et al.(2001b)]{kea01} Keane, J.~V., Tielens,
  A.~G.~G.~M., Boogert, A.~C.~A., Schutte, W.~A., \& Whittet,
  D.~C.~B.\ 2001b, \aap, 376, 254

\bibitem[Kemper et al.(2004)]{kem04} Kemper, F., Vriend, W.~J., \&
  Tielens, A.~G.~G.~M.\ 2004, \apj, 609, 826

\bibitem[Kerkhof et al.(1999)]{ker99} Kerkhof, O., Schutte, W.~A., \&
  Ehrenfreund, P.\ 1999, \aap, 346, 990

\bibitem[Knez et al.(2005)]{kne05} Knez, C., et al.\ 2005, \apjl, 635,
  L145

\bibitem[Lada et al.(1999)]{lad99} Lada, C.~J., Alves, J., \& Lada,
  E.~A.\ 1999, The Physics and Chemistry of the Interstellar Medium,
  161

\bibitem[McClure(2009)]{mcc09} McClure, M.\ 2009, \apjl, 693, L81

\bibitem[McLean et al.(1998)]{mcl98} McLean, I.~S., et al.\ 1998,
  \procspie, 3354, 566

\bibitem[Myers \& Benson(1983)]{mye83} Myers, P.~C., \& Benson,
  P.~J.\ 1983, \apj, 266, 309

\bibitem[{\"O}berg et al.(2007)]{obe07} {\"O}berg, K.~I., Fraser,
  H.~J., Boogert, A.~C.~A., Bisschop, S.~E., Fuchs, G.~W., van
  Dishoeck, E.~F., \& Linnartz, H.\ 2007, \aap, 462, 1187

\bibitem[{\"O}berg et al.(2008)]{obe08} {\"O}berg, K.~I., Boogert,
  A.~C.~A., Pontoppidan, K.~M., Blake, G.~A., Evans, N.~J., Lahuis,
  F., \& van Dishoeck, E.~F.\ 2008, \apj, 678, 1032

\bibitem[Pontoppidan et al.(2003a)]{pon03a} Pontoppidan, K.~M., et
  al.\ 2003a, \aap, 408, 981

\bibitem[Pontoppidan et al.(2003b)]{pon03b} Pontoppidan, K.~M.,
  Dartois, E., van Dishoeck, E.~F., Thi, W.-F., \& d'Hendecourt,
  L.\ 2003b, \aap, 404, L17

\bibitem[Pontoppidan et al.(2008)]{pon08} Pontoppidan, K.~M., et
  al.\ 2008, \apj, 678, 1005

\bibitem[Reach et al.(2009)]{rea09} Reach, W.~T., et al.\ 2009, \apj,
  690, 683

\bibitem[Roche \& Aitken(1984)]{roc84} Roche, P.~F., \& Aitken,
  D.~K.\ 1984, \mnras, 208, 481

\bibitem[Schutte et al.(1993)]{sch93} Schutte, W.~A., Allamandola,
  L.~J., \& Sandford, S.~A.\ 1993, Icarus, 104, 118

\bibitem[Schutte et al.(1999)]{sch99} Schutte, W.~A., et al.\ 1999,
  \aap, 343, 966

\bibitem[Schutte \& Khanna(2003)]{sch03} Schutte, W.~A., \& Khanna,
  R.~K.\ 2003, \aap, 398, 1049

\bibitem[Shu et al.(1987)]{shu87} Shu, F.~H., Adams, F.~C., \& Lizano,
  S.\ 1987, \araa, 25, 23

\bibitem[Skrutskie et al.(2006)]{skr06} Skrutskie, M.~F., et
  al.\ 2006, \aj, 131, 1163

\bibitem[Smith et al.(1989)]{smi89} Smith, R.~G., Sellgren, K., \&
  Tokunaga, A.~T.\ 1989, \apj, 344, 413

\bibitem[Tanaka et al.(1990)]{tan90} Tanaka, M., Sato, S., Nagata, T.,
  \& Yamamoto, T.\ 1990, \apj, 352, 724

\bibitem[Tielens \& Hagen (1982)]{tie82} Tielens, A.~G.~G.~M., \&
  Hagen, W.\ 1982, \aap, 114, 245

\bibitem[Tielens \& Allamandola(1987)]{tie87} Tielens, A.~G.~G.~M., \&
  Allamandola, L.~J.\ 1987, Interstellar Processes, 134, 397

\bibitem[Tielens et al.(1991)]{tie91} Tielens, A.~G.~G.~M., Tokunaga,
  A.~T., Geballe, T.~R., \& Baas, F.\ 1991, \apj, 381, 181

\bibitem[van Breemen et al.(2011)]{bre11} van Breemen, J.~M., et
  al.\ 2011, \aap (in press)

\bibitem[Watanabe \& Kouchi(2002)]{wat02} Watanabe, N., \& Kouchi,
  A.\ 2002, \apjl, 571, L173

\bibitem[Weingartner \& Draine(2001)]{wei01} Weingartner, J.~C., \&
  Draine, B.~T.\ 2001, \apj, 548, 296

\bibitem[Whittet et al.(1983)]{whi83} Whittet, D.~C.~B., Bode, M.~F.,
  Baines, D.~W.~T., Longmore, A.~J., \& Evans, A.\ 1983, \nat, 303,
  218

\bibitem[Whittet et al.(1998)]{whi98} Whittet, D.~C.~B., et al.\ 1998,
  \apjl, 498, L159

\bibitem[Whittet et al.(2001)]{whi01} Whittet, D.~C.~B., Pendleton,
  Y.~J., Gibb, E.~L., Boogert, A.~C.~A., Chiar, J.~E., \& Nummelin,
  A.\ 2001, \apj, 550, 793

\bibitem[Whittet(2003)]{whi03} Whittet, D.~C.~B.\ 2003, Dust in the
  galactic environment, 2nd ed.~ by D.C.B.~Whittet.~Bristol: Institute
  of Physics (IOP) Publishing, 2003 Series in Astronomy and
  Astrophysics, ISBN 0750306246.,

\bibitem[Whittet et al.(2009)]{whi09} Whittet, D.~C.~B., Cook, A.~M.,
  Chiar, J.~E., Pendleton, Y.~J., Shenoy, S.~S., \& Gerakines,
  P.~A.\ 2009, \apj, 695, 94

\bibitem[Young et al.(2004)]{you04} Young, C.~H., et al.\ 2004, \apjs,
  154, 396

\bibitem[Zasowski et al.(2009)]{zas09} Zasowski, G., Kemper, F.,
  Watson, D.~M., Furlan, E., Bohac, C.~J., Hull, C., \& Green,
  J.~D.\ 2009, \apj, 694, 459

\end{thebibliography}
\end{document}